\documentclass[prb]{revtex4}
\usepackage{amssymb}
\usepackage{graphicx}

\begin{document}

\title{Lower Pseudogap Phase: A Spin/Vortex Liquid State}
\author{Zheng-Yu Weng and Xiao-Liang Qi}
\affiliation{Center for Advanced Study, Tsinghua University, Beijing 100084, China}

\begin{abstract}
The pseudogap phase is considered as a new state of matter in the phase
string model of the doped Mott insulator, which is composed of two distinct
regimes known as the \emph{upper} and \emph{lower} pseudogap phases,
respectively. The former corresponds to the formation of spin singlet
pairing whose magnetic characterizations have been recently studied [Phys.
Rev. B \textbf{72}, 104520 (2005)]. The latter as a low-temperature regime
of the pseudogap phase is systematically explored in this work, which is
characterized by the formation of the Cooper pair amplitude and described by
a generalized Gingzburg-Landau theory. Elementary excitation in this phase
is a charge-neutral object carrying spin-$1/2$ and locking with a
supercurrent vortex, known as spinon-vortex composite. Such a lower
pseudogap phase can be regarded as a vortex liquid state due to the presence
of free spinon-vortices. Here thermally excited spinon-vortices destroy the
phase coherence and are responsible for nontrivial Nernst effect and
diamagnetism. The transport entropy and core energy associated with a
spinon-vortex are determined by the spin degrees of freedom. Such a
spontaneous vortex liquid phase can be also considered as a spin liquid with
a finite correlation length and gapped $S=1/2$ excitations,\ where a
resonancelike non-propagating spin mode emerges at the antiferromagnetic
wavevector ($\pi ,\pi $) with a doping-dependent characteristic energy. The
superconducting phase is closely related to the lower pseudogap phase by a
topological transition with spinon-vortices and -antivortices forming bound
pairs and the emergence of fermionic quasiparticles as holon-spinon-vortex
bound objects. A quantitative phase diagram in the parameter space of
doping, temperature, and magnetic field is determined. Comparisons with
experiments are also made.
\end{abstract}

\pacs{74.20.Mn, 74.72.-h, 74.25.Ha, 74.20.De }
\maketitle

\section{Introduction}

Pseudogap phenomenon \cite{timusk,lee1} in the underdoped cuprates is widely
considered to be closely related to the high-$T_{c}$ problem. Any sensible
microscopic theory for the high-$T_{c}$ superconductivity is expected to
include the pseudogap as an integral part of the theory concerning the
high-temperature/energy and short-distance physics. Due to the
unconventional properties manifested \cite{timusk,lee1} in transport, spin
dynamics, optical and single-particle spectroscopies, thermodynamic
properties, etc., experimental measurements in the pseudogap regime have
imposed much\emph{\ stronger }constraints on a potential high-$T_{c}$ theory
than just those in the superconducting state.

Experimentally two distinctive pseudogap regions have been observed\cite%
{timusk,ps1Bi2212,ps2Bi2212,timusk1,ando}. The high-temperature (T) regime,
called the \emph{upper} pseudogap phase (UPP), is marked by the suppression
of uniform magnetic susceptibility \cite{uniform1,uniform2} and the
deviation of dc resistivity from the linear-T behavior \cite%
{uniform2,transp,transp1}, both of which happen below a characteristic
temperature $T_{0}$. The temperature $T_{0}$ is comparable to the
superexchange coupling strength ($\sim 1,000$ K) near the half-filling and
decreases monotonically with doping \cite{uniform1,uniform2,transp,transp1}.
Near the optimal doping, $T_{0}$ is reduced to the same order of magnitude
as the superconducting transition temperature $T_{c}$. The single-particle
pseudogap structure around the wavevector $(\pm \pi ,0)$ and $(0,\pm \pi )$
observed in the ARPES experiments\cite{arpes} also shows a similar doping
dependence. A low-T pseudogap regime, called the \emph{lower} pseudogap
phase (LPP), corresponds to a crossover regime between the UPP and
superconducting (SC) phase, where the low-energy antiferromagnetic (AF)
fluctuations get suppressed as shown in $\mathrm{^{63}Cu}$ NMR experiments
\cite{ps1Bi2212,nmr}, accompanied by the emergence of a \textquotedblleft
high-energy\textquotedblright\ (\textquotedblleft \textrm{41
meV\textquotedblright }) resonancelike peak in the dynamic spin
susceptibility function around the AF wavevector $(\pi ,\pi )$ as revealed
by the neutron measurements \cite{neutron}.

One of the most significant experimental observations in the LPP\ region is
the discovery \cite%
{xu2000,wang2001,wang2002,wang2003,wen2003,wang2005-dia,wang2005} of the
large Nernst signal and residual diamagnetism which strongly support the
presence of vortex excitations \cite{corson} in a wide temperature range
which can be several times higher than $T_{c}$ at low doping. Generally it
is believed that the superconducting phase transition in underdoped
materials is due to the phase disordering effect\cite{emery95} by vortices
as one approaches $T_{c}$ from below. There have been a number of
theoretical proposals \cite%
{emery95,svp2002,tesanovic,lee2,huse2005,anderson2} to explain the existence
of topological vortices above $T_{c}$ in the underdoped regime. However, the
nature of these vortices above $T_{c}$ can vary distinctly from the more
conventional Kosterlitz-Thouless (KT) vortices \cite{emery95,lee2,huse2005}
to exotic vortices with a spin-$1/2$ (spinon) trapped at each core\cite%
{svp2002}.

It would be rather difficult for one to judge the relevance of a particular
microscopic theory simply in terms of some detailed account for individual
experiments, considering huge amounts of existing data available for
interpretation and the lack of consensus on the underlying mechanism after
almost two decades' extensive and intensive investigations. It is thus
particularly meaningful for a potential microscopic theory of high-$T_{c}$
to be worked out extensively and self-consistently to provide a
comprehensive \textquotedblleft road map\textquotedblright . One may then
use the obtained specific \textquotedblleft map\textquotedblright\ and
organizing principles to accommodate and analyze the experimental results
with priority and hierarchy, and to further predict new experiments. Only in
this way may one effectively justify or falsify a particular theoretical
approach based on a broad set of experimental measurements.

The \textquotedblleft road map\textquotedblright\ of the phase string model%
\cite{review-ps2003} is illustrated by a global phase diagram at low doping
in Fig. 1. This model is an effective theory proposed for the cuprate
superconductors based on the $t-J$ model, with a careful treatment of the
spin correlation at various ranges and with \emph{self-consistently }%
incorporating the mutual interplay between the spin and charge degrees of
freedom\cite{review-ps2003,phase-string96,weng98,kou2003,mcs2005,weng2005}.
Such a model describes a doped Mott insulator with spins forming singlet
pairing, known as the bosonic resonating-valence-bond (RVB) pairing, in the
UPP. The magnetic characterizations of this phase have been carefully
examined in Ref. \cite{upp2005} recently. At half-filling, without the
presence of holes, the short-range AF correlations in the UPP will \emph{%
continuously} grow with reducing temperature and eventually becomes AF
long-range ordered (AFLRO) in the ground state (a finite N\'{e}el
temperature will be the consequence of the interlayer coupling). On the
other hand, as doped holes moving in the spin background always generates
the so-called phase string effect\cite{phase-string96}, the spin dynamics
\cite{chen2005} will be drastically reshaped beyond some doping
concentration \cite{kou2003}, in order for holes to further gain kinetic
energy\cite{gu2005} at low-T. In such a finite doping regime, one enters the
LPP [also known as the spontaneous vortex phase (SVP)\cite{svp2002}] at a
low temperature $T_{v}$ and finally ends up in the SC state \cite%
{muthu2002,shaw2003} below $T_{c}$ without encountering an AFLRO.

\begin{figure}[tbp]
\begin{center}
\includegraphics[width=3.5in]{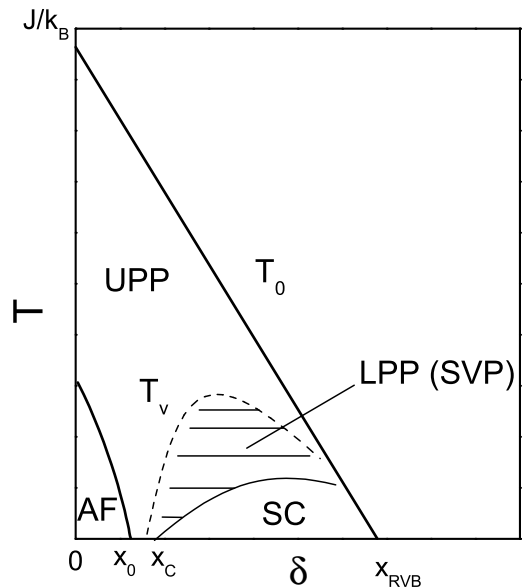}
\end{center}
\caption{The global phase diagram of the phase string model at low doping%
\protect\cite{review-ps2003}. The high-temperature phase is called the upper
pseudogap phase (UPP) which is characterized by the bosonic RVB order
parameter $\Delta ^{s}$ and the temperature $T_{0}.$ The antiferromagnetic
(AF) long-range order state, the lower pseudogap phase (LPP) which is also
called \protect\cite{svp2002} the spontaneous vortex phase (SVP) with a
characteristic temperature $T_{v}$, and the d-wave superconducting phase
(SC) below $T_{c}$, all occur as some forms of low-T instabilities in the
UPP. For example, the AF state corresponds to the spinon condensation on the
RVB background, while the LPP and SC states involve the holon condensation.
The detailed phase diagram is determined by the unique mutual duality gauge
structure \protect\cite{kou2003,mcs2005} which quantitatively governs the
intrinsic competition between the charge and spin degrees of freedom. The
shaded regime of the LPP will be the main focus of this work. }
\label{phasediagram}
\end{figure}

The above global phase diagram with the UPP as the high-T \textquotedblleft
unstable fixed point\textquotedblright\ state of the doped Mott insulator is
spiritually very similar to the proposals made by Anderson \cite%
{anderson1,anderson2} recently, where the LPP and SC phases are considered
as low-energy/temperature instabilities with the spin-charge
\textquotedblleft locking\textquotedblright\ evolving progressively.
Although two approaches look different mathematically, they still share some
important features like the RVB spin pairing in the UPP, vortex liquid state
in the LPP, and vortex-antivortex binding in the SC, as well as the
kinetic-energy-driven and progressive charge-spin recombination in the
latter regimes. The main distinction between the two approaches perhaps lies
in how the UPP is described as the high-energy/temperature starting point.
In a framework based on the Gutzwiller-projected \emph{fermionic} RVB
(f-RVB) mean-field state \cite{lee1,f-rvb}, the short-range AF correlations
are normally intrinsically weak, \emph{before the Gutzwiller projection}, as
compared to those in the \emph{bosonic} RVB state, such that the kinetic
energy of doped holes at the mean-field level is not as severally
frustrated. The advantage of this approach is that the coherent fermion
nodal quasiparticle appears directly in the formalism. By contrast, in the
bosonic RVB state, the short-range AF correlations are intrinsically
associated with the bosonic RVB pairing, which remain strong even \emph{%
before} the Gutzwiller projection is made or in other words, at the
mean-field level, such that the motion of doped holes always gets severely
frustrated in the UPP. Consequently, with decreasing temperature, the
tendency for gaining the kinetic energy becomes the critical driving force
to realize the LPP and SC at a finite doping. Here the fermion nodal
quasiparticle emerges as a coherent one only in the SC state through a more
delicate spin-charge binding process\cite{qp2003}. It is also noted that the
gauge structure describing the fluctuations around the mean-field state,
which is either \textrm{U(1) \cite{u1,lee3},} \textrm{SU(2) \cite{su2} }or
even\textrm{\ Z}$_{2}$ \cite{z2}, arising from the no double occupancy
constraint in the slave-boson approaches\cite{lee1,f-rvb} based on the f-RVB
description, is rather different from the present mutual duality (mutual
Chern-Simons \cite{mcs2005}) gauge structure arising from the phase string
effect in the $t-J$ model\cite{phase-string96}. The latter is crucially
responsible for the global phase diagram shown in Fig. 1, occurring at
proper temperatures and doping regions as compared to the experiments.

In this paper, we will present a systematic and detailed description of the
LPP (the shaded region in Fig. 1) based on the phase string model in the
context of the global phase diagram involving the UPP, SC, and AF phases.
The basic picture is that the pseudogap phase is a new state of matter,
which is a doped Mott insulator with spin-charge separation. In particular,
the LPP is characterized by the holon condensation, described by a
generalized Gingzburg-Landau (GL) theory\cite{muthu2002,svp2002}. The
elementary excitation identified in this phase is a bosonic charge-neutral
object carrying spin-1/2 and vortexlike currents known as a spinon-vortex.
The LPP is composed of thermally excited spinon-vortices which destroy the
SC phase coherence, while the Cooper pair amplitude still remains.
Interesting consequences of this phase will be explored, including the
Nernst effect and diamagnetism. How the LPP is closely related to the SC
phase by a topological transition involving vortex-antivortex binding or
\textquotedblleft spinon confinement\textquotedblright , including the
emergence of coherent\emph{\ fermionic} quasiparticles in the SC phase \cite%
{qp2003},\ will be also discussed.

The remainder of the paper is organized as follows. In Sec. II, the
theoretical framework of the phase string model is presented and the
elementary equations describing the LPP are given. In Sec. III, the physical
properties of the LPP are explored, including the spin liquid behavior,
elementary excitation, phase diagram, the Nernst effect and diamagnetism, as
well as the topological transition to the SC state. Finally, a summary and
conclusions are presented in Sec. IV.

\section{Elementary Equations}

\subsection{Phase string model}

The phase string model is an effective low-energy theory, obtained \cite%
{weng98,weng2005} based on the phase string formulation of the
two-dimensional (2D) $t-J$ model, in which the electron annihilation
operator is decomposed in the following \emph{bosonization} form\cite%
{phase-string96}
\begin{equation}
c_{i\sigma }=h_{i}^{\dagger }b_{i\sigma }e^{i\hat{\Theta}_{i\sigma }}
\label{mutual}
\end{equation}%
where both holon $h_{i\sigma }^{\dagger }$ and spinon $b_{i\sigma }$
operators are bosonic fields, satisfying the no-double-occupancy constraint $%
n_{i}^{h}+\sum_{\sigma }n_{i\sigma }^{b}=1$ with $n_{i}^{h}=h_{i}^{\dagger
}h_{i}$ as the holon number and $n_{i\alpha }^{b}=b_{i\sigma }^{\dagger
}b_{i\sigma }$ as the spinon number. The phase factor $e^{i\hat{\Theta}%
_{i\sigma }}$, which ensures the fermionic statistics of $c_{i\sigma }$, is
defined by $e^{i\hat{\Theta}_{i\sigma }}\equiv e^{i\Theta _{i\sigma
}^{string}}~(\sigma )^{\hat{N}_{h}}(-\sigma )^{i},$ where\ $\hat{N}_{h}$ is
the total holon number operator, $\sigma =\pm 1$, and $(-1)^{i}=\pm 1$ ($%
i\in \mathrm{even}$/\textrm{odd site) }is a staggered sign factor. The phase
$\Theta _{i\sigma }^{string}$ $\equiv \frac{1}{2}\left[ \Phi _{i}^{s}-\Phi
_{i}^{0}-\sigma \Phi _{i}^{h}\right] $ is a nonlocal operator with
\begin{equation}
\Phi _{i}^{s}=\sum_{l(\neq i)}\mbox{Im ln
$(z_i-z_l)$}\left( \sum_{\alpha }\alpha n_{l\alpha }^{b}\right) ~,
\label{phi0}
\end{equation}%
$\Phi _{i}^{0}=\sum_{l(\neq i)}\mbox{Im ln
$(z_i-z_l)$},~$and
\begin{equation}
\Phi _{i}^{h}=\sum_{l(\neq i)}\mbox{Im ln
$(z_i-z_l)$}n_{l}^{h}~,  \label{phih}
\end{equation}%
where $z_{l}=x_{l}+iy_{l}$ is the complex coordinate for a lattice site $l$.

The effective phase string theory is a \emph{bosonic }RVB theory, in which
the spin degrees of freedom are described by the following spinon
Hamiltonian:

\begin{eqnarray}
H_{s} &=&-{J_{s}}\sum_{<ij>\sigma }\left[ \left( e^{i\sigma
A_{ij}^{h}}\right) b_{i\sigma }^{\dagger }b_{j-\sigma }^{\dagger }+\mathrm{%
H.c.}\right] ~  \nonumber \\
&&-\lambda \sum_{i\sigma }\left( b_{i\sigma }^{\dagger }b_{i\sigma }-\frac{%
1-\delta }{2}\right) \text{ \ .}  \label{hspinon}
\end{eqnarray}%
Here the effective coupling constant $J_{s}=J(1-g)\Delta ^{s}/2$, where $%
J\,\ $is the superexchange constant in the original $t-J$ model, $g\simeq
4\delta $ with $\delta $ denoting the doping concentration stands for a
renormalization effect to the superexchange energy $J$, \cite{upp2005} and $%
\Delta ^{s}$ is the bosonic RVB order parameter defined by
\begin{equation}
\Delta ^{s}=\sum_{\sigma }\left\langle e^{-i\sigma A_{ij}^{h}}b_{i\sigma
}b_{j-\sigma }\right\rangle _{\mathrm{NN}}  \label{ds}
\end{equation}%
with the sites $i$ and $j$ belonging to two nearest neighbors (NN). The
Lagrangian multiplier $\lambda $ enforces the constraint $\sum_{\sigma
}\left\langle b_{i\sigma }^{\dagger }b_{i\sigma }\right\rangle =1-\delta $.

The charge degree of freedom is governed by the holon Hamiltonian
\begin{equation}
H_{h}=-\ t_{h}\sum_{<ij>}\left( e^{iA_{ij}^{s}+ieA_{ij}^{e}}\right)
h_{i}^{\dagger }h_{j}+\mathrm{H.c.}  \label{hh}
\end{equation}%
where $t_{h}\sim t$ ($t$ is the hopping integral in the original $t-J$
model) is the renormalized hopping integral for holons, which directly see
the \ external electromagnetic vector potential $A_{ij}^{e}$ with charge $+e$%
.

A peculiar feature of the phase string model is that the spinons and holons
see two \emph{different} gauge fields, $A_{ij}^{h}$ and $A_{ij}^{s}$, with
their strengths constrained to the densities of \emph{holons} and \emph{%
spinons}, respectively. Namely, $A_{ij}^{h}$ and $A_{ij}^{s}$ satisfy the
following gauge-independent topological constraint%
\begin{equation}
\oint\nolimits_{c}d\mathbf{r}\cdot \mathbf{A}^{h}=\pi N^{h}(c)  \label{ah}
\end{equation}%
\begin{equation}
\oint\nolimits_{c}d\mathbf{r}\cdot \mathbf{A}^{s}=\pi \left[ N_{\uparrow
}^{b}(c)-N_{\downarrow }^{b}(c)\right]  \label{as}
\end{equation}%
where $\mathbf{A}^{h}$ and $\mathbf{A}^{s}$ are defined by $A_{ij}^{h}\equiv
\mathbf{r}_{ij}\cdot \mathbf{A}^{h}$ and $A_{ij}^{s}\equiv \mathbf{r}%
_{ij}\cdot \mathbf{A}^{s}$, respectively, with $\mathbf{r}_{ij}$ ($%
\left\vert \mathbf{r}_{ij}\right\vert =a$ is equal to the lattice constant)
a spatial vector connecting two NN sites, $i$ and $j.$ On the right-hand
side (rhs) of the above expressions, $N^{h}(c)$ and $N_{\uparrow ,\downarrow
}^{b}(c)$ denote the numbers of holons and spinons of spin $\uparrow $ or $%
\downarrow $, respectively, which are enclosed by an arbitrary
counter-clockwise-oriented path $c$. Such a topological gauge structure
\emph{entangles} the otherwise decoupled spin and charge degrees of freedom
and is called mutual duality since a spinon sees a holon as a $\pi $ flux
tube and \emph{vice versa}, which has been also formulated in the
path-integral formalism as a mutual Chern-Simons gauge theory\cite{mcs2005}
with both time-reversal and spin rotational symmetries retained.

The phase string model with the mutual duality gauge structure can lead to a
global phase diagram at low-doping as summarized \cite{review-ps2003}in Fig. %
\ref{phasediagram}. It has been shown \cite{upp2005} that as the temperature
is reduced below the characteristic temperature $T_{0}$ of the UPP, the
spins in the background start to form singlet pairs (bosonic RVB pairs) and
condense, which is characterized by the RVB order parameter $\Delta ^{s}$.
The short-range AF correlations simultaneously begin to grow below $T_{0}$,
in contrast to much weaker spin correlations above $T_{0}$.

Several low-temperature phases, including the so-called LPP, which is also
known as the SVP, the d-wave superconducting phase, and the AF phase at low
doping, are all nested below the UPP as shown in Fig. 1 with essentially the
\emph{same }symmetry in the spin sector as characterized by $\Delta ^{s}$.
The competitions between these low-temperature phases are basically decided
by the above mutual-duality (mutual Chern-Simons) gauge structure, rather
than more conventional competitions between different order parameters.

In particular, the LPP corresponds to the occurrence of holon
Bose-condensation $\left\langle h_{i}^{\dagger }\right\rangle \neq 0$
without superconducting phase coherence, whose region is shaded in Fig. \ref%
{phasediagram} and will be the main focus of the following study in this
paper.

\subsection{Superconducting order parameter}

The LPP is the holon condensed phase on a background where spins form the
singlet pairing. In order to understand the corresponding physical
characteristics, it is instructive to examine the superconducting order
parameter $\Delta _{ij}^{\mathrm{SC}}=\sum \sigma \left\langle c_{i\sigma
}c_{j-\sigma }\right\rangle $ below.

Based on the phase string formulation [Eq. (\ref{mutual})], the spinon RVB
paring [Eq. (\ref{ds})], and the holon condensation condition $\left\langle
h_{i}^{\dagger }\right\rangle \neq 0$, the superconducting order parameter
defined at the NN sites can be expressed as \cite{muthu2002} $\Psi _{\mathrm{%
SC}}\equiv \left( \Delta _{ij}^{\mathrm{SC}}\right) _{\mathrm{NN}}=\Delta
_{ij}^{0}\left\langle e^{i\frac{1}{2}(\Phi _{i}^{s}+\Phi
_{j}^{s})}\right\rangle $ where $\Delta _{ij}^{0}\propto \sum_{\sigma
}\left\langle e^{-i\sigma A_{ij}^{h}}b_{i\sigma }b_{j-\sigma }h_{i}^{\dagger
}h_{j}^{\dagger }\right\rangle _{\mathrm{NN}}\simeq \Delta ^{s}\left\langle
h_{i}^{\dagger }\right\rangle \left\langle h_{j}^{\dagger }\right\rangle
\neq 0$. Since in the later GL description we shall treat the charge
condensate in terms of a slowly varying continuous field, $\left\langle
h_{i}^{\dagger }\right\rangle \rightarrow \psi _{h}^{\ast }(\mathbf{r}_{i})$%
, it is convenient to rewrite $\Psi _{\mathrm{SC}}$ in the continuum limit
(without considering the d-wave symmetry of the relative coordinate for
simplicity) as follows
\begin{equation}
\Psi _{\mathrm{SC}}\propto \Delta ^{0}\left\langle e^{i\Phi ^{s}(\mathbf{r}%
)}\right\rangle ~  \label{psisc}
\end{equation}%
where the amplitude
\begin{equation}
\Delta ^{0}=\Delta ^{s}\left( \psi _{h}^{\ast }\right) ^{2}
\label{amplitude}
\end{equation}%
and the phase $\Phi ^{s}(\mathbf{r}_{i})=\Phi _{i}^{s}$ is given by
\begin{equation}
\Phi ^{s}(\mathbf{r})=\int d^{2}\mathbf{r}^{\prime }~\mathrm{Im~ln}\left[
z-z^{\prime }\right] ~\left[ n_{\uparrow }^{b}(\mathbf{r}^{\prime
})-n_{\downarrow }^{b}(\mathbf{r}^{\prime })\right] ~  \label{phis0}
\end{equation}%
with $n_{\sigma }^{b}(\mathbf{r}_{i})\equiv n_{i\sigma }^{b}/a^{2}$. \

\begin{figure}[tbp]
\begin{center}
\includegraphics[width=3.5in]{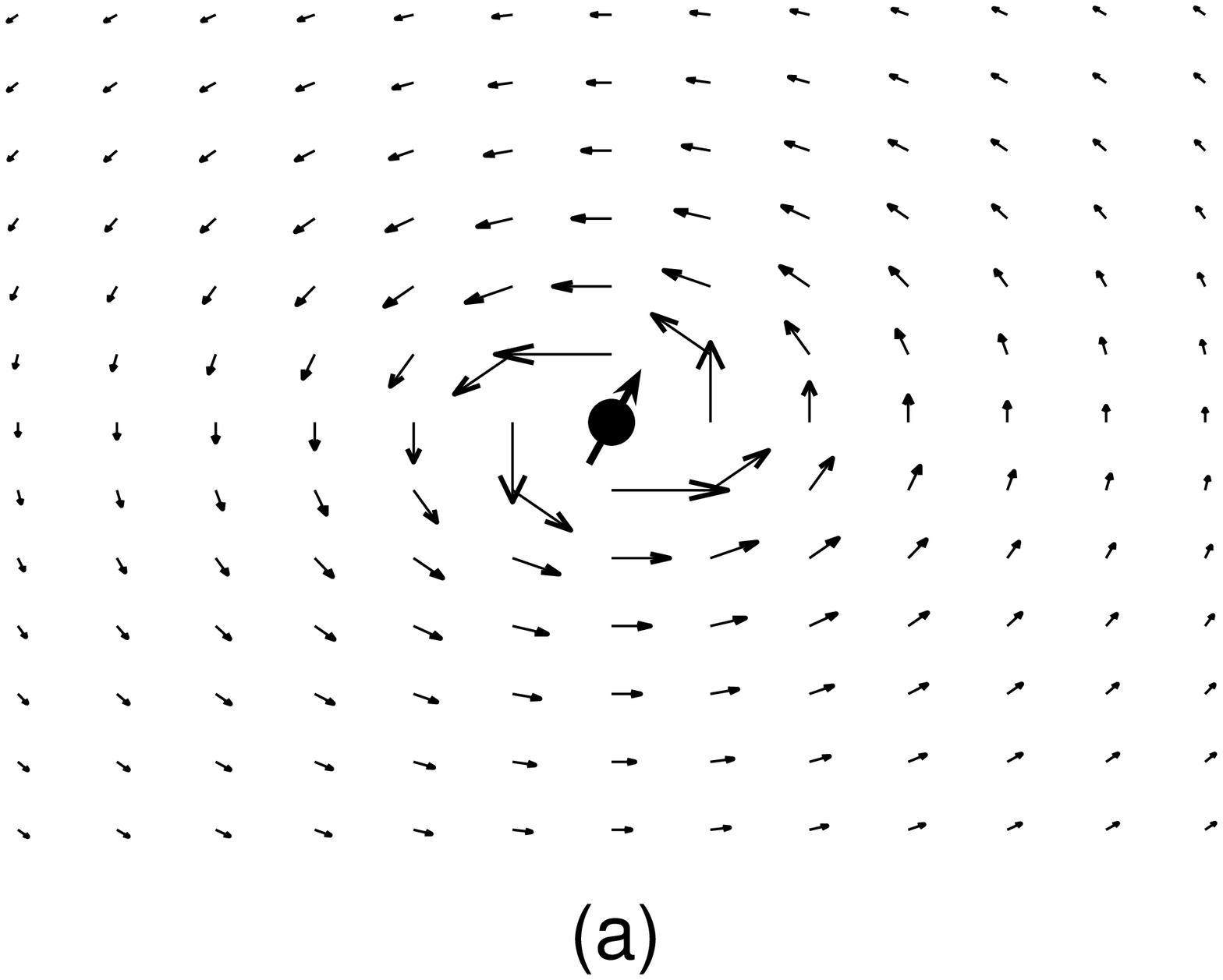} %
\includegraphics[width=3.5in]{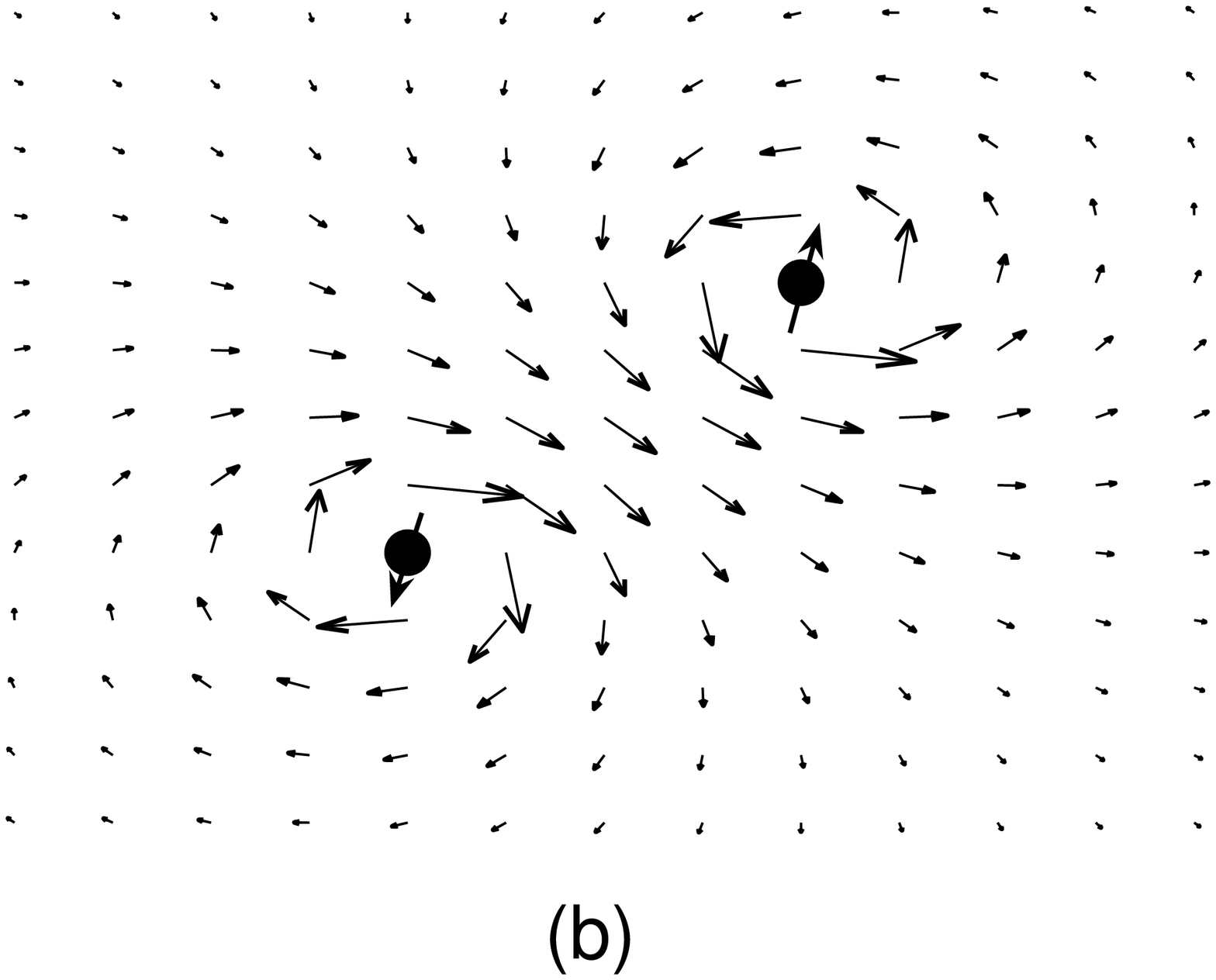}
\end{center}
\par
\caption{ (a) An isolated spinon is always associated with a $2\protect\pi $
vortex in the phase of the superconducting order parameter in the phase
string model. (b) A pair of vortex-antivortex with spinons located at the
cores. }
\label{topo}
\end{figure}

Therefore, in the phase string model the superconducting order parameter $%
\Psi _{\mathrm{SC}}$ has an intrinsic \emph{composite} structure of
amplitude and phase components in Eq. (\ref{psisc}). Here the Cooper pair
amplitude $\Delta ^{0}$ is composed of the RVB pairing $\Delta ^{s}$ and
holon condensate $\psi _{h}^{\ast }$, which is finite at $T\leq T_{v}$ [note
that generally $T_{v}<T_{0}$, as $\psi _{h}\neq 0$ is always underpinned by
the spin singlet paring in the present work (\emph{cf.} Sec. III C)]. Thus
the LPP can be physically regarded as the formation of the Cooper pair
amplitude $\Delta ^{0}\neq 0$.

However, the onset of $\Delta ^{0}$ does not lead to superconductivity
immediately. From Eq. (\ref{phis0}), it is clear that $\Phi ^{s}$ describes $%
2\pi $ phase vortices whose cores are centered at spinons: \emph{i.e.,} $%
\Phi ^{s}\rightarrow \Phi ^{s}\pm 2\pi $ or $\Psi _{\mathrm{SC}}\rightarrow
\Psi _{\mathrm{SC}}e^{\pm 2\pi i}$ each time as the coordinate $\mathbf{r}$
continuously winds around a \emph{spinon} once according to Eq. (\ref{phis0}%
). In other words, a spinon is always associated with a $2\pi $ vortex in $%
\Psi _{\mathrm{SC}}$, known as a \emph{spinon-vortex composite} which is
schematically illustrated in Fig. \ref{topo}(a). A spinon-vortex and
-antivortex pair of a finite range of separation will result in the
cancellation of the phase $\Phi ^{s}$ at a large length scale as shown in
Fig. \ref{topo}(b). Consequently, if there exist excited spinons freed from
the RVB pairing, then $\Phi ^{s}$ is generally disordered such that $%
\left\langle e^{i\Phi ^{s}(\mathbf{r})}\right\rangle =0$ and $\Psi _{\mathrm{%
SC}}=0$, which defines the LPP at $T_{c}<T<T_{v}$. Here the presence of
\emph{free} $2\pi $ vortices in the LPP is a direct manifestation of the
electron \textquotedblleft fractionalization\textquotedblright , \emph{i.e.}%
, the existence of neutral $S=1/2$ spinons in the underlying phase. In the
opposite, the SC phase is determined by the phase coherence $\left\langle
e^{i\Phi ^{s}(\mathbf{r})}\right\rangle \neq 0$ which corresponds to the
\emph{spinon pair confinement} with no single spinon excitations allowed in
the bulk of the sample in the Meissner phase. Thus, the LPP and the
superconducting phase are very closely related, distinct mainly by a
topological transition in the long-distance limit which is to be further
discussed in Sec. III later.

Finally, we point out that a spin $1/2$ bound to the core of a $2\pi $
vortex of the superconducting order parameter is a very general and natural
consequence of doped Mott physics. Namely, with the depleting of superfluid
(hole) density at the core center, due to the no double occupancy
constraint, a spinon trapped inside is energetically most competitive.
Generally there could be also two other ways to create topological
excitations in the superconducting order parameter according to Eqs. (\ref%
{psisc}) and (\ref{amplitude}), which correspond to a $2\pi $-vortex
singularity in the holon condensation order parameter $\psi _{h}(\mathbf{r})$
or the RVB order parameter $\Delta ^{s}$, respectively. However, the former
vortex defined as $\psi _{h}(\mathbf{r})\rightarrow \psi _{h}(\mathbf{r})e^{i%
\mathrm{{Im}\ln }(z-z_{0})}$ leads to a $4\pi $ vortex in $\Psi _{SC}$,
which is energetically not favored. (Nevertheless, such $4\pi $ vortices do
play a role when they are bound to the spinon vortices, as will be shown in
Sec. II C.) The latter one does provide a $2\pi $ vorticity in $\Psi _{SC}$,
but it costs an \emph{expensive} core energy due to the suppression of the
RVB pair order parameter, $\Delta ^{s}\rightarrow 0,$ at the vortex core. By
comparison, $\Delta ^{s}$ can still remain finite in the core of a
spinon-vortex shown above, where the average $\left\langle e^{i\Phi _{s}(%
\mathbf{r})}\right\rangle $ is suppressed to zero within a scale defined by
an intrinsic length scale of the unpaired spinon trapped inside. As to be
shown later, the existence of spinon-vortices in the present theory provides
a natural origin for \textquotedblleft cheap\textquotedblright\ vortices in
the LPP, with the core energy related to a characteristic spin gap scale.

\subsection{Gingzburg-Landau equation}

According to the above definition, the LPP represents a state of matter
characterized by an order parameter field $\psi _{h}(\mathbf{r}),$ on top of
the bosonic RVB condensate, whose microscopic definition is given by
\begin{equation}
\psi _{h}(\mathbf{r})=\left\langle h(\mathbf{r)}\right\rangle  \label{ha}
\end{equation}%
where $h(\mathbf{r)}$ is the continuum version of the bosonic holon matter
field experiencing the Bose condensation. Different from an ordinary
thermodynamic phase like the BCS superconducting state, in which the order
parameter is expressed in terms of the electron operators (\emph{i.e.}, a
pair of the electron creation or annihilation operators), the holon operator
$h(\mathbf{r)}$ cannot be simply expressed based on a local combination of
the electron operators. The corresponding off diagonal long range order
(ODLRO) as suggested by Eq. (\ref{ha}): $\left\langle h^{\dagger }(\mathbf{r)%
}h(\mathbf{r}^{\prime }\mathbf{)}\right\rangle \rightarrow \psi _{h}^{\ast }(%
\mathbf{r})\psi _{h}(\mathbf{r}^{\prime })$ at $\left\vert \mathbf{r-r}%
^{\prime }\right\vert \rightarrow \infty $, must be then a \textquotedblleft
hidden\textquotedblright\ ODLRO, meaning that it involves a nonlocal,
infinite-body order parameter if to be expressed in terms of the electron
operators. More profoundly, since $\psi _{h}(\mathbf{r})$ will couple to an
internal gauge field whose symmetry cannot be spontaneously broken, the
onset of Eq. (\ref{ha}), or more precisely, its amplitude (see below),
should actually represent a crossover rather than a true phase transition.

In the following we introduce an effective GL description of the LPP based
on the order parameter $\psi _{h}$. Let us start with the continuum version
of Eq. (\ref{hh}) which is given by
\begin{equation}
H_{h}={\frac{1}{2m_{h}}}\int d^{2}\mathbf{r}~h^{\dagger }(\mathbf{r})\left(
-i\nabla -\mathbf{A}^{s}-e\mathbf{A}^{e}\right) ^{2}h(\mathbf{r})~
\label{hholon}
\end{equation}%
where $m_{h}$ $=(2t_{h}a^{2})^{-1}$ and $\mathbf{A}^{e}$ is the
electromagnetic vector potential which is seen solely by the holons which
carry charge $+e$. By noting that the holons here are hard-core bosons with
repulsive short-range interaction, one may generally write down the
corresponding GL free energy $F_{h}=\int d^{2}\mathbf{r}$ $f_{h}$ where \cite%
{muthu2002}
\begin{equation}
f_{h}=f_{h}^{0}+\alpha \left\vert \psi _{h}\right\vert ^{2}+\frac{\eta }{2}%
|\psi _{h}|^{4}+\frac{1}{2m_{h}}\psi _{h}^{\ast }\left( -i\nabla -\mathbf{A}%
^{s}-e\mathbf{A}^{e}\right) ^{2}\psi _{h}  \label{fh}
\end{equation}%
with $f_{h}^{0}$ denoting the \textquotedblleft normal
state\textquotedblright\ free energy density. In the absence of fields and
gradients in Eq. (\ref{fh}), the minimum of $f_{h}=f_{h}^{0}+\alpha
\left\vert \psi _{h}\right\vert ^{2}+\frac{\eta }{2}|\psi _{h}|^{4}$ gives
rise to
\[
\left\vert \psi _{h}\right\vert ^{2}=-\frac{\alpha }{\eta }\equiv \rho
_{h}^{0}
\]%
at $\alpha <0$. Here $\rho _{h}^{0}$ represents a bare \textquotedblleft
superfluid\textquotedblright\ density, which is assumed to remain finite
throughout LPP and is reduced to a renormalized $\rho _{h}=\left\vert \psi
_{h}\right\vert ^{2}$ in the presence of $\mathbf{A}^{s}$.

By minimizing $F_{h}$, a nonlinear Schr\"{o}dinger equation or differential
GL equation satisfied by $\psi _{h}(\mathbf{r})$ can be obtained
\begin{equation}
\alpha \psi _{h}+\eta |\psi _{h}|^{2}\psi _{h}+\frac{1}{2m_{h}}\left(
-i\nabla -\mathbf{A}^{s}-e\mathbf{A}^{e}\right) ^{2}\psi _{h}=0  \label{gl0}
\end{equation}%
with the standard \textquotedblleft supercurrent\textquotedblright\ density
given by
\begin{equation}
\mathbf{J}(\mathbf{r})=-\frac{i}{2m_{h}}\left[ \psi _{h}^{\ast }(\mathbf{r}%
)\nabla \psi _{h}(\mathbf{r})-\nabla \psi _{h}^{\ast }(\mathbf{r})\psi _{h}(%
\mathbf{r})\right] -\frac{\mathbf{A}^{s}+e\mathbf{A}^{e}}{m_{h}}\psi
_{h}^{\ast }(\mathbf{r})\psi _{h}(\mathbf{r})~.  \label{j0}
\end{equation}

These equations are similar to an ordinary GL theory describing a charge $+e$
Bose condensate coupled to an external electromagnetic field $\mathbf{A}^{e}$%
, except that $\psi _{h}$ is further coupled to the spin degrees of freedom
through the vector potential $\mathbf{A}^{s}$, whose gauge-independent
definition is given in Eq. (\ref{as}). It means that each isolated spin
(spinon) will register as a $\pm \pi $ flux tube in the nonlinear Schr\"{o}%
dinger equation (\ref{gl0}) to exert frustration effect on the charge
condensate, which reflects the key influence of the spin degrees of freedom
on the charge condensate in the phase string model.

The GL equations, Eqs. (\ref{gl0}) and (\ref{j0}), are gauge invariant under
the transformation: $\psi _{h}\rightarrow \psi _{h}e^{i\varphi }$ and $%
\mathbf{A}^{s}\rightarrow \mathbf{A}^{s}+\nabla \varphi $ where $\varphi $
is an arbitrary single-valued function. Especially, one may also construct a
\emph{large} gauge transformation $\varphi =\sum\nolimits_{i}\varphi _{i}$,
with
\begin{equation}
\oint_{c_{i}}d\mathbf{r}\cdot \nabla \varphi _{i}=\pm 2\pi  \label{phi2pi}
\end{equation}%
where a loop $c_{i}$ encircles a $\pm 2\pi $ vortex center $i$ which
coincides with a spinon position. This procedure is allowed because in the
underlying microscopic theory a holon cannot stay at the sites of spinons
due to the no double occupancy constraint and $\psi _{h}$ should thus vanish
at a spinon position where a $\pm \pi $ flux tube is bound to according to
Eq. (\ref{as}) ($\mathbf{A}^{s}$ diverges there). Through such kind of large
gauge transformation, the rhs. of Eq. (\ref{as}) (if one spinon is enclosed
by $c$) may be properly transformed by a minus sign as
\begin{equation}
\oint\nolimits_{c}d\mathbf{r}\cdot \mathbf{A}^{s}=\pm \pi \rightarrow
\oint\nolimits_{c}d\mathbf{r}\cdot \mathbf{A}^{s}=\mp \pi  \label{as1}
\end{equation}%
which means the sign of vorticity around a spinon is no longer directly tied
to the spin index $\sigma =\uparrow $, $\downarrow $. In this way the theory
is expressed in an explicitly spin-rotational invariant form, where the sign
of the $\pi $ fluxoid carried by each spin excitation (spinon) can be
determined by minimizing the free energy instead.

The above large gauge transformation can be also visualized based on the
superconducting order parameter $\Psi _{\mathrm{SC}}$ defined in Eq. (\ref%
{psisc}). Evidently, $\Phi ^{s}$ is related to $\mathbf{A}^{s}$ ($\nabla
\Phi ^{s}=2\mathbf{A}^{s}$), satisfying the gauge invariance under the gauge
transformation: $\psi _{h}\rightarrow \psi _{h}e^{i\varphi }$, $\mathbf{A}%
^{s}\rightarrow \mathbf{A}^{s}+\nabla \varphi $, and $\Phi ^{s}\rightarrow
\Phi ^{s}+2\varphi $. If one makes \emph{large }gauge transformations like
Eq. (\ref{phi2pi}), then the vorticity signs of the $2\pi $ vortices in $%
\Phi ^{s}$ can be changed such that $n_{\uparrow }^{b}(\mathbf{r})$ and $%
n_{\downarrow }^{b}(\mathbf{r})$ generally should be replaced by
\begin{equation}
\Phi ^{s}(\mathbf{r})=\int d^{2}\mathbf{r}^{\prime }~\mathrm{Im~ln}\left[
z-z^{\prime }\right] ~\left[ n_{-}^{b}(\mathbf{r}^{\prime })-n_{+}^{b}(%
\mathbf{r}^{\prime })\right] ~,  \label{phis1}
\end{equation}%
where $n_{\pm }^{b}(\mathbf{r})$ denotes excited spinon number with
vorticity $\pm 2\pi $ no longer be directly tied to the spin index (the
definition of $\pm $ sign for the vorticity is such that $+$ vortices are
commensurate with the magnetic field applied along the perpendicular
direction of the 2D plane, \emph{cf.} Sec. III B). In order words, although $%
\left( \psi _{h}^{\ast }\right) ^{2}$ only contributes $4\pi $ vortices to $%
\Psi _{\mathrm{SC}}$, it can also change the sign of $2\pi $ vortices in $%
\Phi ^{s}$ if the vortex cores of these $4\pi $ vortices \emph{properly}
coincide with the spinons (note that the higher vorticities like $6\pi $ per
spinon as the result of the above transformation will cost too much energy
and are thus not considered here).

In summary, the LPP phase is described by a generalized GL description of
the holon condensate coupled with unbound spinon-vortices, the dynamics of
which is governed by the spinon Hamiltonian (\ref{hspinon}). In the next
section, the physical properties of the LPP phase will be explored based on
such an effective description.

\section{Physical Properties of Lower Pseudogap Phase}

Let us first focus on the spin degrees of freedom below, which will play an
important role in the LPP.

\subsection{Spin degrees of freedom: A spin liquid}

Let us recall that the spinon mean-field Hamiltonian (\ref{hspinon}) can be
diagonalized by the following Bogoliubov transformation \cite{weng98}
\begin{equation}
b_{i\sigma }=\sum_{m}w_{m\sigma }\left( \mathbf{r}_{i}\right) \left[
u_{m}\gamma _{m\sigma }-v_{m}\gamma _{m-\sigma }^{\dagger }\right] \text{ }
\label{bogo}
\end{equation}%
with $u_{m}=1/\sqrt{2}(\lambda /E_{m}+1)^{1/2}$\ and $v_{m}=1/\sqrt{2}%
(\lambda /E_{m}-1)^{1/2}\mathrm{sgn}(\xi _{m})$. The diagonalized
Hamiltonian is written as
\begin{equation}
H_{s}=\sum_{m\sigma }E_{m\sigma }\gamma _{m\sigma }^{\dagger }\gamma
_{m\sigma }+\mathrm{const.}  \label{hs}
\end{equation}%
where the spinon spectrum $E_{m}=\sqrt{\lambda ^{2}-\xi _{m}^{2}}$ and $\xi
_{m}$ is the eigenvalue of a tight-binding equation
\begin{equation}
\xi _{m}w_{m\sigma }\left( \mathbf{r}_{i}\right)
=-J_{s}\sum_{j=NN(i)}e^{-i\sigma A_{ji}^{h}}w_{m\sigma }(\mathbf{r}_{j})
\label{bdg}
\end{equation}%
Here the gauge field $A_{ij}^{h}$ is constrained to the hole number
according to Eq. (\ref{ah}), which incorporates the most nontrivial doping
effect on the spin degrees of freedom in the phase string model. In the LPP
when holons are Bose condensed, $A_{ij}^{h}$ may be treated as describing a
uniform flux of strength $\delta \pi $ per plaquette.

A spinon excitation created by $\gamma _{m\sigma }^{\dagger }$ with an
eigen-energy $E_{m}$ has no direct gauge coupling with the external
electromagnetic field $\mathbf{A}^{e}$ in the mean-field spinon Hamiltonian $%
H_{s}$ [Eq. (\ref{hs})] and is thus a charge-neutral, $S=1/2$ object. Here a
magnetic field can still couple to spinons via the Zeeman energy, $-$ $\mu _{%
\mathrm{B}}B$ $\sum_{i\sigma }\sigma n_{i\sigma }^{b}$. Adding this term to $%
H_{s}$ will lead to a modified spinon spectrum in Eq. (\ref{hs})

\begin{equation}
E_{m\sigma }=E_{m}-\sigma \mu _{\mathrm{B}}B  \label{spinonsp}
\end{equation}%
In obtaining Eq. (\ref{spinonsp}) one need to note that
\begin{equation}
\sum_{i\sigma }\sigma n_{i\sigma }^{b}=\sum_{m\sigma }\sigma n_{m\sigma
}^{\gamma }\   \label{as2}
\end{equation}%
with using $u_{m}^{2}-v_{m}^{2}=1$ and $\sum\limits_{i}\left\vert w_{m\sigma
}(\mathbf{r}_{i})\right\vert ^{2}=1$, where the excited spinon number $%
n_{m\sigma }^{\gamma }$ is defined by
\begin{equation}
n_{m\sigma }^{\gamma }\equiv \gamma _{m\sigma }^{\dagger }\gamma _{m\sigma }.
\label{ng}
\end{equation}

In the LPP, by treating $A_{ij}^{h}$ as a vector potential for a \emph{%
uniform} flux perpendicular to the 2D plane with the field strength $B^{h}=%
\frac{\pi \delta }{a^{2}},$ the spinon wave function $w_{m\sigma }\left(
\mathbf{r}\right) $ determined by Eq. (\ref{bdg}) is characterized by a
wavepacket of a typical size of the cyclotron length scale $a_{c}=1/\sqrt{%
B^{h}},$ $\emph{i.e.,}$
\begin{equation}
a_{c}=\frac{a}{\sqrt{\pi \delta }}\text{.}  \label{ac}
\end{equation}%
The RVB pairing of spins described by $H_{s}$ in Eq. (\ref{hs}) has the
following mean-field form in the ground state \cite{weng2005}%
\[
|\mathrm{RVB}\rangle =C\exp \left( \sum_{ij}W_{ij}b_{i\uparrow }^{\dagger
}b_{j\downarrow }^{\dagger }\right) |\mathrm{0}\rangle
\]%
where the RVB amplitude $\left\vert W_{ij}\right\vert =\left\vert \sum_{m}%
\frac{v_{m}}{u_{m}}w_{m\sigma }^{\ast }\left( \mathbf{r}_{i}\right)
w_{m\sigma }\left( \mathbf{r}_{j}\right) \right\vert $ behaves as follows
\[
\left\vert W_{ij}\right\vert \propto e^{-\left\vert \mathbf{r}_{i}-\mathbf{r}%
_{j}\right\vert ^{2}/2\xi _{\mathrm{s}}^{2}}
\]%
at $\left\vert \mathbf{r}_{i}-\mathbf{r}_{j}\right\vert \gg a,$ for $ij$
belonging to different sublattices. Here the characteristic size for the RVB
pair is given by \cite{weng2005}
\begin{equation}
\xi _{\mathrm{s}}=a\sqrt{\frac{2}{\pi \delta }}  \label{xis}
\end{equation}%
which also decides the equal-time spin-spin correlation length. Note that $%
\xi _{\mathrm{s}}$ diverges at small $\delta $ where an AFLRO emerges.
Namely, the spin sector is described by \emph{a spin liquid} at finite
doping, which can be continuously connected to the AFLRO state as the doping
concentration is reduced to zero.

\begin{figure}[tbp]
\begin{center}
\includegraphics[width=3.5in]{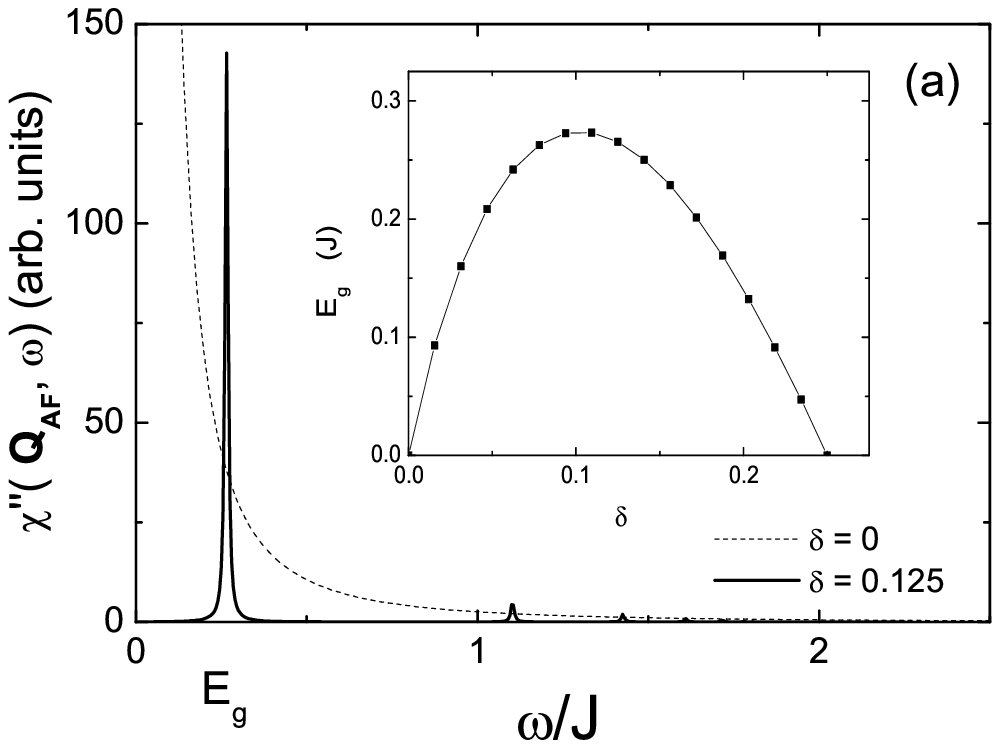} 
\includegraphics[width=3.5in]{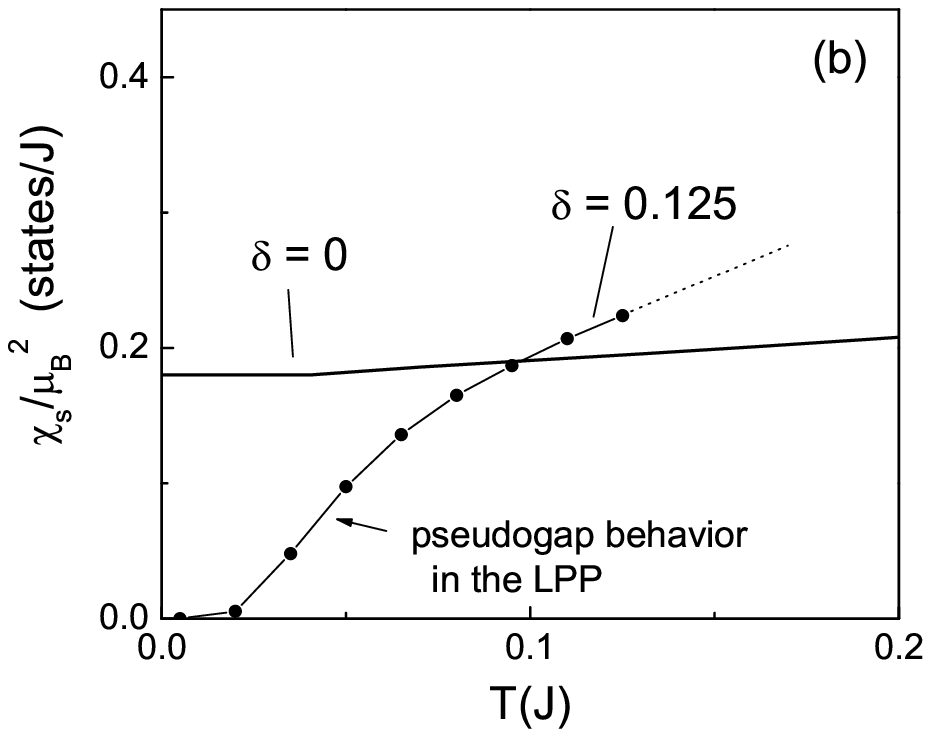}
\end{center}
\caption{Spin liquid behavior in the LPP at $\protect\delta =0.125$ as
compared to the half-filling case with an AFLRO. (a) Dynamic spin
susceptibility at momentum $\mathbf{Q}_{\mathrm{AF}}=(\protect\pi ,\protect%
\pi )$. Inset: the evolution of the resonancelike peak energy $E_{g}$ as a
function of doping concentration. (b) Uniform spin susceptibility versus the
temperature. }
\label{spinon}
\end{figure}

Thus, although there exist a lot of spins at low doping ($1-\delta $ per
site), the majority of them form short-range RVB pairs of a typical size $%
\xi _{\mathrm{s}}$ in the holon condensation phase. According to Eq. (\ref%
{as}), then the contribution of these RVB paired spins to $\mathbf{A}^{s}$
will be effectively cancelled out, and the charge condensate described by $%
\psi _{h}$ is essentially decoupled from the RVB background at a scale
larger than $\xi _{\mathrm{s}}$. This is also self-consistent with the fact
that $\xi _{\mathrm{s}}$ is compatible with the average holon-holon distance
which sets a minimal length scale for the holon condensate. Only the excited
spinons will then effectively contribute to $\mathbf{A}^{s}$ in the GL
equation by noting that
\begin{equation}
\sum_{\sigma }\sigma \left\langle n_{i\sigma }^{b}\right\rangle
=\sum_{m\sigma }\sigma \left\vert w_{m\sigma }(\mathbf{r}_{i})\right\vert
^{2}\left\langle n_{m\sigma }^{\gamma }\right\rangle \   \label{sdensity}
\end{equation}%
based on the Bogoliubov transformation (\ref{bogo}). In obtaining Eq. (\ref%
{sdensity}), one has used the fact that $\left\vert w_{m\sigma }(\mathbf{r}%
)\right\vert ^{2}$ is independent of $\sigma $. Here the length scale
determined by $\left\vert w_{m\sigma }(\mathbf{r}_{i})\right\vert ^{2}$,
\emph{i.e., }the \textquotedblleft cyclotron\textquotedblright\ length $%
a_{c} $ in Eq. (\ref{ac}) sets a natural spinon size which is within the
\textquotedblleft coarse graining\textquotedblright\ length $\xi _{\mathrm{s}%
}$ as $a_{c}=\xi _{\mathrm{s}}/\sqrt{2}$ $\lesssim \xi _{\mathrm{s}}$. Thus
an excited spinon can still maintain its quantum integrity in the GL
equation description.

Eq. (\ref{hs}) describes the spinon excitations as neutral spin-$1/2$
excitations in the RVB background, labelled by $m$ and spin index $\sigma $.
Fig. \ref{spinon}(a) shows the corresponding dynamic spin susceptibility
function at the AF wavevector $\mathbf{Q}_{\mathrm{AF}}=(\pi ,\pi )$ for $%
\delta =0$ and $\delta =0.125$, respectively, in which a resonancelike peak
naturally emerges at finite doping. In the inset, such a resonance energy $%
E_{g}$ as a function of doping is presented (the bare superexchange coupling
$J$ used in Ref. \cite{chen2005} should be replaced by a renormalized one $%
J_{\mathrm{eff}}$ in the expression of $J_{s}$ as discussed in Ref. \cite%
{upp2005}). Note that more precisely, the competing AF state can drive $%
E_{g} $ vanish at $x_{c}\sim 0.04$ in a fashion of $E_{g}\sim \sqrt{\delta
-x_{c}}J$,\cite{kou2003} instead of $E_{g}\sim \delta J$ at $\delta
\rightarrow 0$ as shown in Fig. \ref{spinon}(a), which is beyond the
mean-field description presented here. In Fig. \ref{spinon}(b), the uniform
spin susceptibility\cite{upp2005} is also shown for both $\delta =0$ and $%
\delta =0.125,$ and the later one clearly exhibits a pseudogap behavior at
low temperature. These typical magnetic properties evolving from the undoped
AF phase to the LPP are quite consistent with the experimental measurements
in the high-$T_{c}$ cuprates.

\subsection{Spin-charge entanglement in the LPP: Spinon-vortex composite}

As discussed above, the spin sector in the LPP can be characterized as a
spin liquid state with charge neutral, spin 1/2 spinons as its elementary
excitations. Here one finds essentially the same symmetry for the spin
degrees of freedom in both the UPP and LPP, which are described by the
bosonic RVB order parameter $\Delta ^{s}$. What really differentiates the
UPP and LPP lies in the charge sector by the additional holon condensation $%
\psi _{h}\neq 0$ in the LPP. As it turns out, this distinction in the charge
degree of freedom then \emph{feedbacks} to the spin sector to affect the
spin excitations, resulting in a spin gap behavior in the LPP as represented
in Fig. \ref{spinon} in the previous section.

In the following we shall discuss a more striking feature due to the mutual
entanglement between the spin and charge degrees of freedom based on the GL
equations. Namely, a spinon excitation will always induce a current vortex
in the LPP to form a spinon-current-vortex composite, which is consistent
with the $2\pi $ vortices previously identified in the phase of the
superconducting order parameter in Sec. II B [Fig. 2(a)].

\begin{figure}[tbp]
\begin{center}
\includegraphics[width=3.5in]{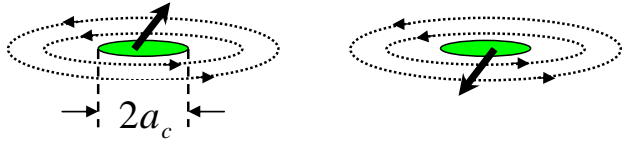}
\end{center}
\caption{A neutral spin-$1/2$ bosonic object, spinon, is always locking with
a current vortex as an elementary excitation in the LPP, known as a
spinon-vortex composite. The vortex core size is decided by the quantum
length scale of the spinon, $a_{c}$, with its vorticity sign independent of
the spin index.}
\label{fig_cvortex}
\end{figure}

By writing%
\begin{equation}
\psi _{h}=\sqrt{\rho _{h}}e^{i\phi _{h}}  \label{psih}
\end{equation}%
a London equation for the supercurrent based on Eq. (\ref{j0}) (neglecting
the spatial dependence of $\rho _{h}$) is given by%
\begin{equation}
\mathbf{J}(\mathbf{r})=\frac{\rho _{h}}{m_{h}}\left[ \nabla \phi _{h}-%
\mathbf{A}^{s}-e\mathbf{A}^{e}\right]  \label{j2}
\end{equation}%
Since each unpaired spinon will contribute to $\oint\nolimits_{c}d\mathbf{r}%
\cdot \mathbf{A}^{s}=\pm \pi $ in terms Eq. (\ref{as}) if the loop $c$
encloses such a spinon of a typical radius of $a_{c}$, a \emph{minimal }%
supercurrent vortex centered around it is then given by
\begin{equation}
\oint\nolimits_{c}d\mathbf{r}\cdot \mathbf{J}(\mathbf{r})=\pm \pi \frac{\rho
_{h}}{m_{h}}  \label{cvortex}
\end{equation}%
according to Eq. (\ref{j2}). This spinon-vortex composite is illustrated in
Fig. \ref{fig_cvortex}, where the \textquotedblleft
cyclotron\textquotedblright\ length $a_{c}$ serves as a natural radius for
the core size of a spinon-vortex.

As discussed in Sec. II C, a large gauge transformation in $\phi _{h}$ can
result in a binding of a $2\pi $ vortex to a spinon to change the sign of
vorticity it carries. Consequently, a spinon excitation is always a
composite of a neutral spin-$1/2$ locking with a charge current vortex in
the LPP, but the vorticity and spin index are not necessarily the same. Such
a spinon-vortex binding is a key consequence of holon condensation and play
an essential role in understanding the LPP physics. Due to the presence of
unconfined spinon-vortices, the LPP is a phase-frustrated BEC of holon and
is more precisely characterized by the finite amplitude of $\psi _{h}$, $%
\sqrt{\rho _{h}}$.

By requiring that
\[
\left\langle \oint\nolimits_{c}d\mathbf{r}\cdot \mathbf{J}(\mathbf{r}%
)\right\rangle =0
\]%
and omitting \emph{independent} $2\pi $ vortices in $\nabla \phi _{h}$
(which are presumably important only at temperatures close to $T_{v})$, the
following \textquotedblleft neutrality\textquotedblright\ condition is then
obtained%
\begin{eqnarray}
\oint\nolimits_{c}d\mathbf{r}\cdot \mathbf{A}^{e} &=&\frac{1}{e}%
\oint\nolimits_{c}d\mathbf{r}\cdot \left( \nabla \phi _{h}-\mathbf{A}%
^{s}\right)  \nonumber \\
&=&\phi _{0}\left[ N_{+}(c)-N_{-}(c)\right]  \label{neutral}
\end{eqnarray}%
where $\phi _{0}\equiv hc/2e$ and $N_{\pm }(c)$ denotes the number of
spinon-vortices (antivortices) enclosed in an arbitrary loop $c$. It shows
that a perpendicular magnetic field will generally polarize the numbers of $%
\pm \pi $ spinon-vortices (the sign $\pm $ is defined according to Eq. (\ref%
{neutral})) in the LPP.

Thus, spinon excitations in the LPP are no longer \textquotedblleft
free\textquotedblright\ ones solely determined by $H_{s}$ in Eq. (\ref{hs}).
They will actually acquire a long-range logarithmic interaction due to the
spinon-vortex effect. Indeed, substituting Eq. (\ref{psih}) into Eq. (\ref%
{fh}) and carrying out the area integration, one finally obtains
\begin{equation}
{F}_{h}=\int \int d^{2}\mathbf{r}_{1}d^{2}\mathbf{r}_{2}\left[
\sum\nolimits_{\alpha }\alpha n_{\alpha }^{b}(\mathbf{r}_{1})\right] V(%
\mathbf{r}_{12})\left[ \sum\nolimits_{\beta }\beta n_{\beta }^{b}(\mathbf{r}%
_{2})\right] +\mathrm{const.}  \label{hv}
\end{equation}%
in which $\alpha ,\beta =\pm $ refer to the signs of vorticities carried by
spinons and
\begin{equation}
V(\mathbf{r}_{12})=-\frac{\pi \rho _{h}}{{4m_{h}}}\ln \frac{|\mathbf{r}_{1}-%
\mathbf{r}_{2}|}{r_{c}}
\end{equation}%
with $r_{c}\sim a$. Nevertheless, in the LPP where a substantial number of
spinon-vortices is present, the long-range force in Eq. (\ref{hv}) will be
screened such that spinon-vortices are not bound together, as opposed to the
lower-temperature superconducting regime where all excited spinons form
vortices-antivortices bound pairs (\emph{cf. }Sec. III F). Namely the LPP
remains an electron fractionalized state before it undergoes the
superconducting phase transition at $T_{c}$.

However, in the LPP, the short-range interaction in Eq. (\ref{hv}) remains
unscreened. Each excited spinon at a state labelled by $m$ has a spatial
extension determined by $\left\vert w_{m}(\mathbf{r})\right\vert ^{2}$ as
mentioned before. With involving the charge current, a symmetric profile of $%
\left\vert w_{m}(\mathbf{r})\right\vert ^{2}$ in the symmetric gauge is
energetically most favorable. Then a pair of spinon excitations at the same $%
m$ with opposite vorticities have essentially no contribution from Eq. (\ref%
{hv}) due to the cancellation, but one with the same vorticities will cost a
big repulsive core energy $\sim \frac{\pi \rho _{h}}{{m_{h}}}$ in $F_{h}$.
Such a short-range interaction effect will be further considered later when
we discuss the diamagnetism in the LPP.

\subsection{Phase diagram}

The phase diagram of the doped Mott insulator can be drawn by combining the
description of the LPP phase here and the previous work on the UPP phase\cite%
{upp2005}.

\begin{figure}[tbp]
\begin{center}
\includegraphics[width=3.5in]{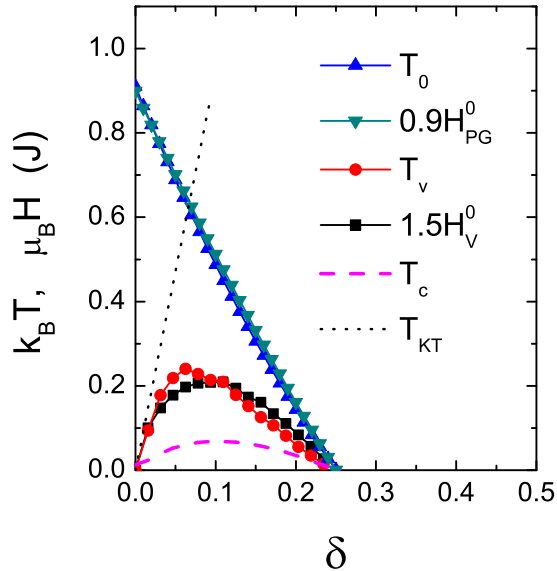}
\end{center}
\caption{The characteristic temperature and magnetic field scales which
determine the phase diagram of the upper and lower pseudogap and
superconducting phases based on the phase string model. Here $T_{0}$ and $H_{%
\mathrm{PG}}^{0}\equiv $ $H_{\mathrm{PG}}(T=0)$ for the UPP, $T_{v}$ and $%
H_{v}^{0}\equiv H_{v}^{{}}(T=0)$ for the LPP, and $T_{c}$ for the SC phase.}
\label{phase}
\end{figure}

\emph{The upper pseudogap phase.---}The phase string model holds for the
whole bosonic RVB regime underpinned by $\Delta ^{s}\neq 0,$ which defines
the UPP as illustrated in Fig. 1. The characteristic temperature $T_{0}$ and
magnetic field $H_{\mathrm{PG}}^{0}$ (at $T=0)$ for the UPP at the boundary $%
\Delta ^{s}=0$ have been previously determined in Ref.\ \cite{upp2005},
which are consistent with the experimental measurements\cite%
{uniform1,uniform2,transp,transp1}. Such temperature and magnetic field
scales are replotted as functions of doping with $x_{\mathrm{RVB}}=0.25$ in
Fig. \ref{phase} (filled triangles).
\begin{figure}[tbp]
\begin{center}
\includegraphics[width=3.5in]{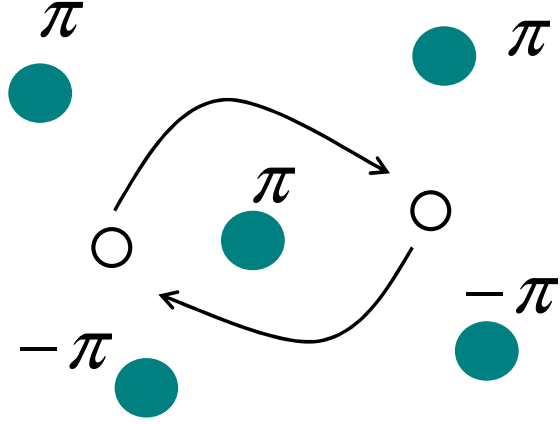}
\end{center}
\caption{Two holons (open circles) exchanging positions can pick up a minus
sign if an unpaired spinon (grey circle) is enclosed in between, which
carries a $\pm \protect\pi $ flux tube due to the mutual duality gauge
structure in the phase string model. So the phase coherence of bosonic
holons can be effectively destroyed to prevent the Bose condensation if
there is a sufficient number of free spinon excitations in the singlet spin
background. }
\label{holonexchange}
\end{figure}

\emph{The lower pseudogap phase.---}Now let us consider the phase boundary
of the LPP, which is set by vanishing $\psi _{h}$ at $T_{v}$. Note that
without $\mathbf{A}^{s}$ this would be a simple 2D hard-core boson problem
according to Eq. (\ref{hholon}), and the KT temperature for the holon
condensation is given by%
\begin{equation}
T_{\mathrm{KT}}=\pi \delta \left( 2a^{2}m_{h}\right) ^{-1}  \label{tkt}
\end{equation}%
which has been previously used\cite{svp2002} as an upper bound temperature
for the LPP at low doping, shown in Fig. 5 by a dotted line (with $t_{h}=3J$%
).

However, the frustration effect of $\mathbf{A}^{s}$ on the holon
condensation can be very important. The spinon-vortex density is determined
by the density (per site) of spinon Bogoliubov quasiparticles as $%
n_{v}=\sum_{m\sigma }\left\langle n_{m\sigma }^{\gamma }\right\rangle /N$.
Due to the opening up of a spin gap $E_{g}$ in the holon condensation phase (%
\emph{i.e., }the LPP), $n_{v}$ is exponentially small for $T\ll E_{g}$. With
increasing temperature, $n_{v}$ will monotonically increase until reaching
the maximal number $n_{v}^{\mathrm{max}}=1-\delta $ at $T=T_{0}$ where all
the RVB pairs break up. 

Since each free spinon carries a $\pi $ fluxoid as perceived by the holons,
the quantum phase coherence among bosonic holons will be violently
interrupted if on average there is an excited (unpaired) spinon sitting
between each two neighbored holons (as illustrated by Fig. \ref%
{holonexchange}), where an exchange between a pair of holons can gain a
minus sign in wavefunction. In other words, the holon condensation must
break down when the vortex density $n_{v}$ is comparable to holon density $%
\delta $, far less than $n_{v}^{\mathrm{max}}$ at low doping. Such a
consideration provides an estimate of the upper bound for $T_{v}$ as
\begin{equation}
n_{v}=\frac{1}{N}\sum_{m\sigma }\frac{1}{e^{E_{m\sigma }/k_{B}T_{v}}-1}%
=\delta  \label{tv}
\end{equation}%
$T_{v}$ determined by Eq. (\ref{tv}) is plotted in Fig. \ref{phase} by
filled circles (with $E_{m\sigma }$ determined by the self-consistent
mean-field theory\cite{upp2005} of the UPP as $\psi _{h}\rightarrow 0$). Eq.
(\ref{tv}) can be also understood based on the \textquotedblleft core
touching\textquotedblright\ picture of spontaneously excited spinon vortices
in the LPP\cite{remark1}. Note that the average distance between excited
spinons may be defined by $l_{s}\equiv 2a/\sqrt{\pi n_{v}}.$ Since the
characteristic core size of a spinon-vortex is $2a_{c}$ as shown in Fig. \ref%
{fig_cvortex}, then one expects that the \textquotedblleft
supercurrents\textquotedblright\ carried by spinon-vortices are totally
destroyed when the sample is fully packed by the \textquotedblleft
cores\textquotedblright\ with $l_{s}=2a_{c}$ which results in Eq. (\ref{tv}%
). Here it is emphasized that there is no simple mean-field phase transition
for the LPP due to the peculiar mutual duality gauge structure mediated by $%
\mathbf{A}^{s}$ and $\mathbf{A}^{h}$. Furthermore, $2\pi $ vortices of $\psi
_{h}$, independent of spinon-vortices, will be also generally present near $%
T_{v}$ as $\rho _{h}\rightarrow 0$. All of these fluctuation effects can
affect the crossover from the LPP to UPP in details.

\begin{figure}[tbp]
\begin{center}
\includegraphics[width=3.5in]{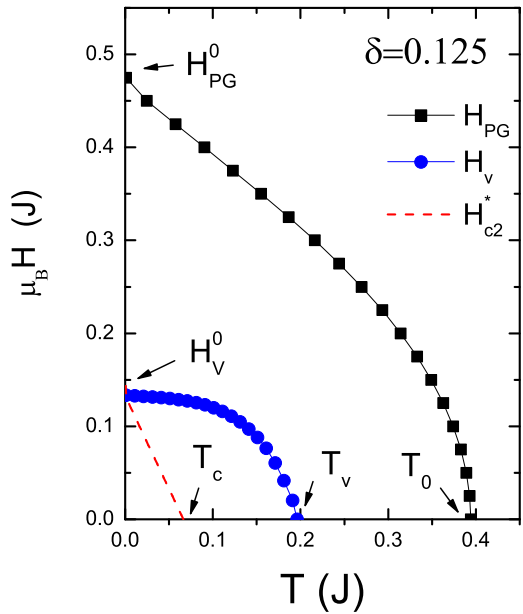}
\end{center}
\caption{The magnetic field -- temperature phase diagram of the pseudogap
phases at doping $\protect\delta =0.125$.}
\label{HTPhD}
\end{figure}

Similar to the UPP case\cite{upp2005}, an external magnetic field will break
up some RVB pairs through the Zeeman effect to create more excited spinons
at a given temperature. By considering the Zeeman effect on the energy
spectrum $E_{m\sigma }$ in Eq. (\ref{spinonsp}), the magnetic field
dependence of $T_{v}=T_{v}(H)$ can be further obtained from Eq. (\ref{tv}).
Or conversely, for each $T<T_{v}$ there is a characteristic field $H_{v}(T)$
at which the LPP phase is destructed. $H_{v}^{0}\equiv H_{v}$($T=0)$
determined this way is shown in Fig. \ref{phase} (filled squares). Roughly
we find $k_{\mathrm{B}}T_{v}\sim 1.5\mu _{\mathrm{B}}H_{v}^{0}$ according to
Fig. \ref{phase}, or $H_{v}^{0}/T_{v}\sim 0.99$ (T/\textrm{K})$\ $which is
slightly larger than the experimental ratio $\sim 0.62$ (\textrm{T/K) }%
determined by the Nernst measurement\cite{wang2005} in the LSCO compound.
Furthermore, $H_{v}$ as a function of temperature is further shown in Fig. %
\ref{HTPhD} at doping concentration $\delta =0.125$ (in the same figure, the
critical field $H_{\mathrm{PG}}(T)$ for the UPP is also shown, which
terminates at $T=T_{0}$).

\emph{Superconducting phase.---}The superconducting phase transition (the
dashed curve in Fig. \ref{phase}) is given by \cite{shaw2003}
\begin{equation}
T_{c}\simeq \frac{E_{g}}{4k_{\mathrm{B}}}  \label{tc}
\end{equation}%
which is determined by the characteristic energy $E_{g}$ of the low-lying
spin excitation---the so-called resonancelike peak energy (\emph{cf.} Fig. %
\ref{spinon}). Such a transition is non-BCS-like and is represented by a
topological transition related to the phase coherence of the superconducting
order parameter, which will be further discussed in Sec. III F.

In the mixed state below $T_{c}$, by including the Zeeman energy according
to Eq.(\ref{spinonsp}), $E_{g}$ is reduced to $E_{g}^{\ast }=E_{g}(B)-2\mu _{%
\mathrm{B}}B$ such that one can estimate $T_{c}(B)$ by using a simple
relation $T_{c}(B)\sim E_{g}^{\ast }/4k_{\mathrm{B}}.$ Then in turn one may
define $H_{c2}^{\ast }\equiv B(T_{c}),$ which\ is shown in Fig. \ref{HTPhD}
by a dashed curve (here a self-consistent mean-field solution of $%
E_{g}^{\ast }$ in the LPP, in the presence of the Zeeman term, is
numerically determined). Note that $H_{c2}^{\ast }$ so defined will vanish
at $T_{c},$ resembling the conventional $H_{c2}$ in a BCS superconductor.
However, since free spinon-vortices are generally present at $H>$ $%
H_{c2}^{\ast }$, $H_{c2}^{\ast }$ is a crossover field which no longer has
the same meaning as $H_{c2}$ in the conventional BCS superconductor. Roughly
speaking, the Abrikosov magnetic vortices (their total number is
proportional to the magnetic field) are expected to be present below $%
H_{c2}^{\ast }$ where the additional spinon-vortices generated by the Zeeman
effect, with both signs in the vorticity, are still loosely paired, while
the vortices unbinding occurs above $H_{c2}^{\ast }$. So $H_{c2}^{\ast }$
defines a crossover between two types of vortex regime. Of course, if the
Abrikosov vortex lattice\ are melt below $H_{c2}^{\ast }$, the distinction
between the two vortex liquid regimes should be further blurred at $%
H_{c2}^{\ast }.$ At $T=0$, $H_{c2}^{\ast }(0)$ obtained by setting $%
E_{g}^{\ast }=0$ is found to be numerically quite close to $H_{v}^{0}$ at $%
\delta =0.125$ (Fig. \ref{HTPhD}), although there is no physical reason for
them to be the same. The numerical result also shows that $\mu _{\mathrm{B}%
}H_{c2}^{\ast }(0)\simeq E_{g}(B=0,T=0)/2$, or
\begin{equation}
H_{c2}^{\ast }(0)\simeq 3T_{c}\left( \frac{\mathrm{T}}{\mathrm{K}}\right)
\label{hc2}
\end{equation}%
according to Eq. (\ref{tc}) which is in quantitatively agreement with the
critical magnetic field extrapolated from the specific heat measurement \cite%
{wen2005} in the SC phase by fitting the Volovik$\sqrt{B}$ term of a d-wave
superconductor in the presence of a perpendicular magnetic field.
\begin{figure}[t]
\begin{center}
\includegraphics[width=3.5in]{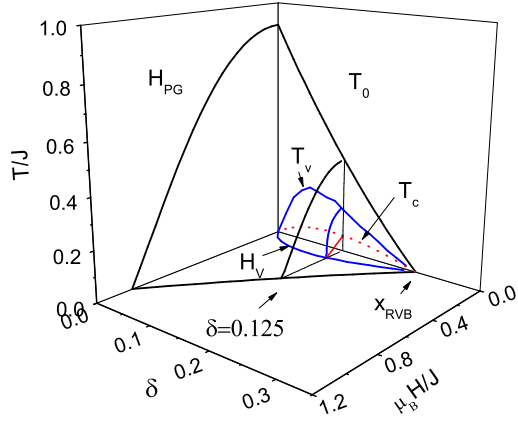}
\end{center}
\caption{{}The phase diagram of the pseudogap regimes in the
three-dimensional space of magnetic field, doping, and temperature.}
\label{phase3D}
\end{figure}

Therefore, the LPP is distinct from the UPP by an additional charge
condensation, and the former is always nested inside the latter because, in
order to realize the condensation in the charge sector, the excited
spinon-vortex density should be restricted to avoid too strong the
frustrations on the former---each unpaired (excited) spinon always exerts
destructive phase interference on the charge condensate in a form of
carrying a $\pi $ flux tube. Here one clearly sees how the mutual duality of
the gauge structure plays a crucial role in fixing the phase boundary.
Finally, a three-dimensional phase diagram of the UPP, LPP and SC phase in
the parameter of magnetic field, doping concentration, and the temperature
determined by the above methods is summarized in Fig. \ref{phase3D}.

\subsection{Nernst effect}

In the absence of the integrity of fermionic quasiparticles (see Sec. III
F), the transport properties in the LPP are mainly associated with the
motion of spinon-vortices. This kind of transport is drastically different
from the usual normal state transport contributed by quasiparticles, which
more resembles that in the flux-flow phase of a BCS superconductor. But
there are some crucial distinctions between the latter and the LPP. In a
conventional type II superconductor, the vortex core energy is usually very
large compared to the temperature $T\sim T_{c}$, which means that the
thermal fluctuations of vortices can be ignored and nearly all the vortices
are generated by an external magnetic field such that $n_{v}=B/\phi _{0}$.
On the opposite, in the present LPP, the spinon-vortices as quantum objects
can be created thermally, which means $n_{v}\gg B/\phi _{0}$ in weak field
case. It is those \emph{spontaneous} vortices that will decide the transport
phenomenon in such a region.

Let us start with the charge current given by the London-like equation in
Eq. (\ref{j2}). By using the steady current condition%
\begin{equation}
\partial _{t}\mathbf{J}=0  \label{j1}
\end{equation}%
and the definition of the electric field $\mathbf{E}=-\partial _{t}\mathbf{A}%
^{e}$ in the transverse gauge, one can finally obtain the following basic
relation
\begin{equation}
\mathbf{E}=\mathbf{\hat{z}}\times \phi _{0}\left( n_{v}^{+}\mathbf{v}%
_{+}-n_{v}^{-}\mathbf{v}_{-}\right) \text{ \ }  \label{ey}
\end{equation}%
where $n_{v}^{\pm }$ denotes the density of spinon-vortices and
-antivortices with drifting velocity $\mathbf{v}_{\pm }$ along a direction
perpendicular to the electric field in the plane. As illustrated by Fig. \ref%
{SchematicNernst}(a), the electric field and the drifting of vortices and
antivortices must be balanced according to Eq. (\ref{ey}) in order to avoid
the system being accelerated indefinitely ($\partial _{t}\mathbf{J}\neq 0$).
So the applied electric field will drive the vortices and antivortices
moving along a perpendicular direction with opposite velocities: $\mathbf{v}%
_{+}=-\mathbf{v}_{-}$ if the vortices and antivortices are not polarized by
the external magnetic field, \emph{i.e., }$n_{v}^{+}$ $\mathbf{=}$ $%
n_{v}^{-} $. Equation (\ref{ey}) is a basic relation for the
spinon-vortex-related transport phenomenon in the LPP.

\begin{figure}[tbp]
\begin{center}
\includegraphics[width=5in]{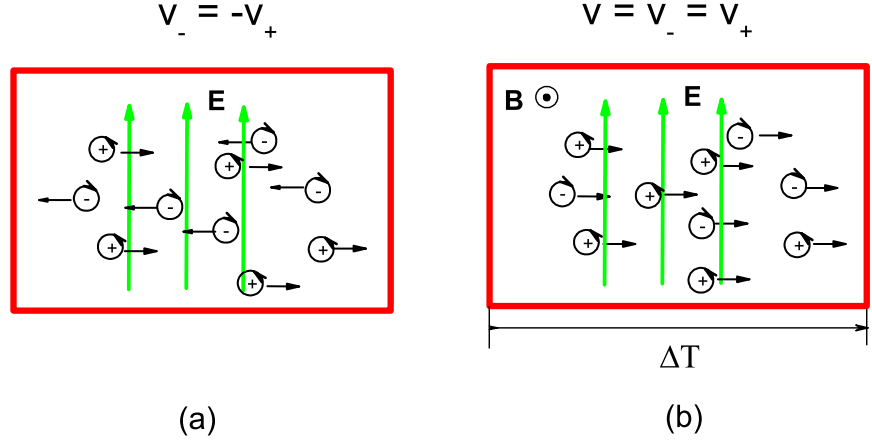}
\end{center}
\caption{Schematic picture of (a) the flux flowing under an electric field
in the LPP, which can lead to a flux flow resistivity as well as the spin
Hall effect, and (b) the flux flowing under a temperature gradient $\protect%
\nabla T$, which must be balanced by an electric field $E$ and thus leads to
the Nernst effect. }
\label{SchematicNernst}
\end{figure}

The Nernst signal corresponds to the electric field measured along the $\hat{%
y}$-direction when spinon-vortices and -antivortices are both driven by a
temperature gradient in the \emph{same} direction along the $\hat{x}$%
-direction. Such a case is shown in Fig. \ref{SchematicNernst}(b), where the
spinon-vortices and -antivortices move along the\emph{\ }$\hat{x}$-direction
with the same velocity $\mathbf{v}_{+}=\mathbf{v}_{-}=\mathbf{v}$ in a
steady state. To have a finite $\mathbf{E}$ in terms of Eq. (\ref{ey}), then
the vortex density $n_{v}^{\pm }$ has to be polarized by the external
magnetic field $\mathbf{B=}B\mathbf{\mathbf{\hat{z}}}$ according to the
\textquotedblleft neutrality\textquotedblright\ condition (\ref{neutral}):

\begin{equation}
B=\phi _{0}\left( n_{v}^{+}-n_{v}^{-}\right)  \label{b}
\end{equation}%
such that
\[
\mathbf{E=\mathbf{\hat{z}}\times }\phi _{0}\left( n_{v}^{+}-n_{v}^{-}\right)
\mathbf{v}=\mathbf{B}\times \mathbf{v}
\]

Define the Nernst signal $e_{N}$ by%
\begin{equation}
e_{N}=\frac{E_{y}}{-\nabla _{x}T}\text{ \ .}  \label{nuv0}
\end{equation}%
Suppose $s_{\phi }$ is the \emph{transport} entropy carried by a spinon
vortex and $\eta _{s}$ is its viscosity such that the drift velocity $%
\mathbf{v}^{s}$ can be decided by $s_{\phi }\nabla T=-\eta _{s}\mathbf{v.}$
Then one has%
\begin{equation}
e_{N}=B\frac{s_{\phi }}{\eta _{s}}  \label{nu2}
\end{equation}%
\bigskip

On the other hand, in the absence of the temperature gradient, a charge
current can also drive a transverse motion of spinon-vortices and
-antivortices along \emph{opposite} directions, i.e., $\mathbf{v}_{\pm }=\pm
\mathbf{v},$ such that an electric field is generated along the current
direction according to Eq. (\ref{ey}), leading to a finite resistivity due
to the presence of free vortices, which is given by\cite{she2005}

\begin{equation}
\rho =\frac{n_{v}}{\eta _{s}}\phi _{0}^{2}  \label{rho}
\end{equation}%
where
\[
n_{v}\equiv n_{v}^{+}+n_{v}^{-}\text{ \ }
\]%
denotes the total density of vortices and antivortices. This formula is
familiar in the vortex flow regime of a conventional superconductor. Then,
by eliminating $\eta _{s}$ which is harder to calculate, one obtains%
\begin{eqnarray}
\alpha _{xy} &\equiv &\frac{e_{N}}{\rho }  \nonumber \\
&=&\frac{Bs_{\phi }}{\phi _{0}^{2}n_{v}}  \label{alpha}
\end{eqnarray}%
where $\alpha _{xy}$ is the quantity introduced in Ref. \cite{wang2001}. It
is noted that in the vortex phase of a conventional superconductor, one has $%
B=\phi _{0}n_{v}^{+}=\phi _{0}n_{v}$ with $n_{v}^{-}=0$ such that Eq. (\ref%
{alpha}) reduces to $\alpha _{xy}\phi _{0}=s_{\phi }$ from which the
transport entropy of a vortex can be obtained directly (e.g., Ref.\cite%
{wang2002}). But in the LPP, $\alpha _{xy}\phi _{0}=s_{\phi }(B/\phi
_{0}n_{v})<s_{\phi }$, which means the transport entropy is underestimated
by $\alpha _{xy}\phi _{0}$ here.
\begin{figure}[tbp]
\begin{center}
\includegraphics[width=3.5in]{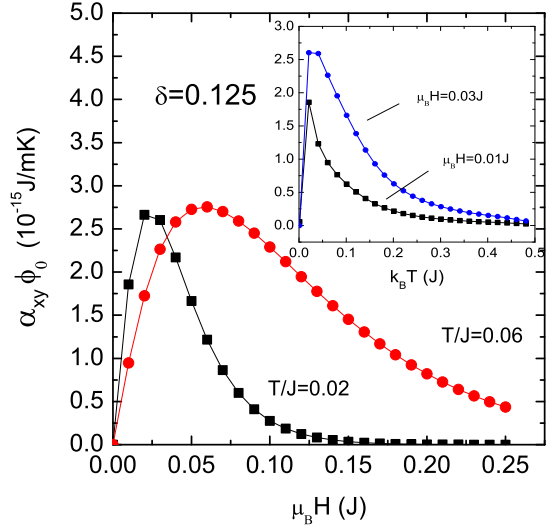}
\end{center}
\caption{{}The quantity $\protect\alpha _{xy}\protect\phi _{0}/d\equiv e_{N}%
\protect\phi _{0}/\protect\rho d$, which is related to the Nernst signal
without involving the viscosity coefficient, is shown as a function of the
magnetic field and temperature (the inset) at a given doping. Here $d=7.7%
\mathring{A}$ is the distance between two $\mathrm{CuO}_{2}$ layers.}
\label{figalpha}
\end{figure}

So far all the derivations are similar to the flux flow region in a
conventional superconductor except that the total vortex number $n_{v}$ in
Eq. (\ref{alpha}) generally is no longer simply proportional to the applied
magnetic field $B$, which remains finite even in the limit $B\rightarrow 0$
to account for thermally activated spinon-vortices. What really makes the
Nernst transport unique in the present theory is that the transport entropy $%
s_{\phi }$ here is due to the spin degree of freedom associated with a
spinon-vortex, instead of a normal core in a conventional BCS
superconductor. Let us recall that the core size of a spinon-vortex is
determined by the \textquotedblleft cyclotron\textquotedblright\ length
scale $a_{c}$. If one neglects the \textquotedblleft
configuration\textquotedblright\ entropy associated with the spatial
location of a spinon-vortex, then the main \textquotedblleft transport\
entropy\textquotedblright\ as carried by a spinon-vortex is due to its free $%
S=1/2$ moment which gives rise to

\begin{equation}
s_{\phi }=k_{\mathrm{B}}\left\{ \ln \left[ 2\cosh \left( \beta \mu _{\mathrm{%
B}}B\right) \right] -\beta \mu _{\mathrm{B}}B\tanh \left( \beta \mu _{%
\mathrm{B}}B\right) \right\} \text{ \ .}  \label{sphi}
\end{equation}%
Here the interaction between spinons in different vortex cores are ignored,
which will in general lead to an additional suppression of the average
entropy per vortex. Under such an approximation, $s_{\phi }\rightarrow 1/2$
for $T\gg \mu _{B}H$ and $s_{\phi }\rightarrow 0$ for $T\ll \mu _{B}H$.
Despite of the saturation of $s_{\phi }$ at high temperature, the quantity $%
\alpha _{xy}$ defined in Eq. (\ref{alpha}) decreases at high temperature due
to the quick proliferation of the total vortex number. The temperature and
magnetic field dependence of $\alpha _{xy}\phi _{0}/d$ is shown in Fig. \ref%
{figalpha}, in which $d=7.7\mathring{A}$ is the distance between two $%
\mathrm{CuO}_{2}$ layers. The magnitude of such a quantity is quite
comparable to the experimental data \cite{wang2002}, implying that the
transport entropy due to the free moment in a spinon-vortex is capable to
produce the Nernst signal observed experimentally. However, at strong
magnetic field or high temperature, $\alpha _{xy}$ decays slower than that
seen in the real systems. That is because here we have only considered
single rigid spinon-vortices without taking into account the interaction
between spinon-vortices as well as the fluctuations of vortices, which
become very important as the LPP will eventually collapse at $H_{v}$ and $%
T_{v}$ as discussed before.

Finally, we briefly mention a unique prediction related to the spinon-vortex
motion driven by an external electric field. As shown in Fig. \ref%
{SchematicNernst}(a), vortices can be driven by an in-plane electric field
to move along the transverse direction. Since each vortex carries a free
moment, if these moments are partially polarized, say, by an external
magnetic field via the Zeeman effect, then a spin Hall current can be
generated along the vortex motion direction. The spin Hall conductivity has
been recently obtained \cite{she2005} as follows:

\begin{equation}
\sigma _{H}^{s}=\frac{\hbar \chi _{s}}{g\mu _{B}}\left( \frac{B}{n_{v}\phi
_{0}}\right) ^{2}  \label{sigmh}
\end{equation}%
which only depends on the intrinsic properties of the system like the
uniform spin susceptibility $\chi _{s},$ with the electron $g$-factor $%
g\simeq 2$. It is important to note that the external magnetic field $B$
applied perpendicular to the 2D plane reduces the spin rotational symmetry
of the system to the conservation of the $S^{z}$ component only, satisfying $%
\frac{\partial S^{z}}{\partial t}+\nabla \cdot \mathbf{J}^{s}=0.$ Thus the
polarized spin current $\mathbf{J}^{s}$ is still conserved and remains
dissipationless as the \emph{current} of its carriers -- vortices is \emph{%
dissipationless} in the LPP. In contrast, the charge current remains
dissipative with the resistivity given by Eq. (\ref{rho}). The
dissipationless spin Hall effect in the LPP is therefore an important
prediction of the phase string model. For more details one is referred to
Ref. \cite{she2005}.

\subsection{Diamagnetism}

In an ordinary superconductor, the diamagnetism can be well described in
terms of the GL free energy. In such a description, the vortex phase below $%
T_{c}$, as induced by magnetic field, exhibits strong residual diamagnetism,
continuously evolving from a perfect Meissner state and eventually vanishing
as the magnetic field $H\rightarrow H_{c2}$ or temperature $T\rightarrow
T_{c}$ when the Abrikosov vortices become closely packing (\textquotedblleft
core touching\textquotedblright ). The diamagnetism here is directly
connected to the energy cost for creating a vortex which carries swirling
supercurrents and is subject to long-range interactions with other vortices,
where the total number of vortices is usually proportional to the applied
magnetic field by $n_{v}=B/\phi _{0}$.

The present LPP is a vortex liquid state \emph{above} $T_{c}$, and it would
be natural for one to conjecture that a substantial diamagnetism still
persist. However, like the Nernst effect discussed above, due to the
involvement of \emph{spontaneous} vortices as well as the spin degrees of
freedom, the theoretical description of the diamagnetism will be also
modified drastically. In the present of thermal vortices, a weak magnetic
field will mainly polarize the numbers of vortices and antivortices, instead
of creating new vortices, and thus the diamagnetism due to the energy cost
of the vortex creation will diminish quickly. On the other hand, if there is
a dense population of thermal vortices, then the configuration entropy will
sensitively depend on the imbalance of the numbers of vortices and
antivortices, due to the \textquotedblleft hard-core\textquotedblright\ like
repulsion between two vortices of the same sign of vorticity as discussed
before, which will cause an additional diamagnetism response.

Let us note that the total free energy for the LPP is given by%
\begin{equation}
F=F_{h}+F_{s}+\int d^{2}\mathbf{r}\frac{B^{2}}{8\pi }  \label{F}
\end{equation}%
where $B$ is the magnetic field perpendicular to the plane and $F_{s}$ is
the free energy for the spin part: $F_{s}=-\beta ^{-1}\ln Tr\left( e^{-\beta
H_{s}}\right) .$ The magnetization $M=(B-H)/4\pi $ is determined by%
\begin{equation}
M=-\frac{1}{\Omega }\left( \frac{\partial F_{h}}{\partial B}+\frac{\partial
F_{s}}{\partial B}\right) _{T}  \label{M}
\end{equation}%
where $\Omega $ denotes the 2D area of the system.

\begin{figure}[tbp]
\begin{center}
\includegraphics[width=3.5in]{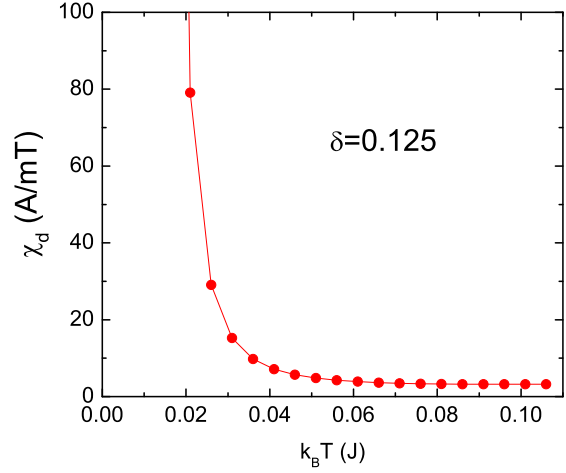}
\end{center}
\caption{{}The diamagnetic susceptibility versus temperature in the LPP.}
\label{chid}
\end{figure}

As discussed before, Eq. (\ref{M}) is composed of two kinds of sources for
diamagnetism. The first one is similar to that in the conventional vortex
phase of a BCS superconductor contributed by supercurrents associated with
vortices. But the vortex liquid phase \emph{above} $T_{c}$ is represented by
the presence of a sufficient number of thermally excited free vortices that
destroy the superconducting phase coherence. These spontaneous free vortices
are already present in the absence of an external magnetic field
perpendicular to the 2D plane, due to the screening of the long-range
(logarithmic) interaction among vortices [Eq.(\ref{hv})]. So in the vortex
liquid phase, the free energy $F$, to the leading order approximation,
should only depend on the total number $n_{v}$ of vortices and antivortices,
each with the same renormalized self-energy. On the other hand, when the
applied magnetic field is weak enough as compared to the total number per
unit area of thermally-excited vortices multiplying the flux quantum $\phi
_{0}$ (i.e., $B\ll n_{v}\phi _{0}$), it can only partially polarize the
numbers of $+$ and $-$ vortices by Eq.\ (\ref{b}): $\Delta n_{v}=B/\phi _{0}$%
, while the total vortex number $n_{v}$ remains unchanged. Without an
explicit dependence of $F$ on the polarization $\Delta n_{v}$ of vortices
and antivortices that is proportional to the magnetic field, one would find
a vanishing diamagnetic response to the weak magnetic field.

Another origin of diamagnetism is related to the entropy part of the free
energy, as already pointed out above. In the following we provide an
estimate of this effect as a \emph{lower} bound of the diamagnetism at weak
field limit. Recall that we have previously discussed the quantum mechanical
nature of the core of a spinon-vortex: the trapped spinon does a cyclotron
motion within the core according to $H_{s}$, and its Hilbert space is
restricted to the lowest Landau level (LLL) at low temperatures ($\gtrsim
T_{c}$). The above-mentioned short-range interactions will then dramatically
affect the entropy of spinon-vortices: two vortices with the same vorticity
will cost a large, unscreened energy if they stay in the same real-space
(symmetric) cyclotron orbit labelled by $m,$ whereas two with opposite
vortices approximately do not experience any substantial short-range
interaction even when they are in the same real-space orbit. Then we can
approximately write down the total spinon-vortex configuration entropy as%
\begin{equation}
S_{s}=\Omega k_{\mathrm{B}}\sum_{\alpha }\left[ n_{\alpha }^{s}\ln \left(
\frac{G}{n_{\alpha }^{s}}-1\right) -g\ln \left( 1-\frac{n_{\alpha }^{s}}{G}%
\right) \right]  \label{ss}
\end{equation}%
where $N$ is the number of lattice sites and $N_{\alpha }^{s}=n_{\alpha
}^{s}N$ is the spinon number with spin index $\alpha $. $G=\delta /2\times
2=\delta $ denotes the degeneracy of the LLL per unit area, including the
spin index, and the condition that two same vortices cannot occupy the same
state labelled by $m$ has been used with assumption that at $T\gtrsim T_{c}$
mainly the LLL is involved.

When the magnetic field is in the range of $B\ll n_{v}\phi _{0}$, one can
reasonably assume $\delta B=\phi _{0}\left( \delta n_{v}^{+}-\delta
n_{v}^{-}\right) =2\phi _{0}\delta n_{v}^{+}=-2\phi _{0}\delta n_{v}^{-}$
since $\delta n_{v}=\delta n_{v}^{+}+\delta n_{v}^{-}\simeq 0$. \ Then, the
magnetization determined by the entropy $S_{s}$ is given as follows%
\begin{eqnarray*}
M_{s} &\simeq &\frac{T}{\Omega }\left( \frac{\partial S_{s}}{\partial B}%
\right) _{n_{v}}=\frac{T}{\Omega (2\phi _{0})}\left( \frac{\partial S_{s}}{%
\partial n_{v}^{+}}\right) _{n_{v}} \\
&=&\frac{k_{\mathrm{B}}T}{2\phi _{0}}\ln \left( \frac{G-n_{+}^{s}}{%
G-n_{-}^{s}}\frac{n_{-}^{s}}{n_{+}^{s}}\right)
\end{eqnarray*}

At weak field limit, the rhs of the above expression may be expanded to the
first order of $B$ as
\[
M_{s}=-\chi _{d}^{0}B,\text{ or }M_{s}=\chi _{d}H
\]%
with
\begin{equation}
\chi _{d}=-\frac{\chi _{0}^{d}}{1+4\pi \chi _{0}^{d}}  \label{chid1}
\end{equation}%
where $\chi _{d}^{0}=\frac{2Gk_{\mathrm{B}}T}{(2G-n_{v})n_{v}\phi _{0}^{2}}$%
. \ In Fig. \ref{chid}, $\chi _{d}^{0}$ is shown as a function of the
temperature at $\delta =0.125,$ whose magnitude is comparable to the
experimental result\cite{wang2005-dia} at $H=1$ \textrm{T. }Note that at $%
T=0 $ one has $n_{v}=0$ and $\chi _{d}^{0}\rightarrow -\infty $ such that $%
M_{s}\rightarrow -\left( 1/4\pi \right) H$, where the full Meissner state is
recovered. In reality, one has $n_{v}=0$ below $T_{c}$ and thus expects to
see a divergence in $\chi _{d}$ there. Near $T_{c}$ the correction from the
superconducting phase fluctuations may be important to result in some
nonlinear singular field dependence\cite{wang2005-dia} of $M_{s}$ which is
not included in Eq. (\ref{chid1}). Furthermore, $\chi _{d}$ reaches a
minimum but does not vanish at $T_{v}$ where $n_{v}\sim \delta $ because the
rigid spinon-vortices are considered here without taking into account of the
collapse of spinon-vortices close to $T_{v}.$ Finally, as remarked before,
here we have only focused on the weak field limit $B\ll n_{v}\phi _{0}$.
With increasing the magnetic field, more spinon-vortices will be generated
and nonlinear effect at strong magnetic field will emerge which is explored
quantitatively elsewhere\cite{qi2006}.

\subsection{Superconducting phase transition}

With $\Delta ^{0}\neq 0$ in the LPP, the superconducting state can be
finally realized through the phase coherence\cite{muthu2002}%
\begin{equation}
\left\langle e^{i\Phi ^{s}(\mathbf{r})}\right\rangle \neq 0  \label{pc}
\end{equation}%
in which the spinon-vortex composites must form vortex-antivortex pairs [%
\emph{cf.} Fig. \ref{topo}(b)] in an analog with the KT transition in the XY
model.

The corresponding dynamics is decided by Eq. (\ref{hv}). It is similar to
the XY model, except that $\mathbf{A}^{s}$ introduces $\pi $ instead $2\pi $
vortices and the vortex cores are attached to spinons which have their own
dynamics as governed by $H_{s}$ with an intrinsic quantum length scale $%
a_{c} $. Consequently the superconducting phase transition is not of a
BCS-type second-order one as described by an ordinary GL theory --- here $%
\psi _{h}$ or more precisely the amplitude $\sqrt{\rho _{h}},$ is in general
not expected to vanish at $T_{c}$, especially in the underdoped regime. It
is rather a phase coherence transition involving spinon-vortices
binding/unbinding which directly affects $e^{i\Phi ^{s}(\mathbf{r})}$. The
detailed renormalization group analysis \cite{shaw2003} leads to the $T_{c}$
relation (\ref{tc}) which connects the phase coherence temperature with the
spin resonancelike energy $E_{g}$ in a quantitative agreement with
experimental measurements\cite{bourges98}.

The superconducting phase coherence (\ref{pc}) implies that spinons are
\emph{confined} in the bulk where a single spinon-vortex excitation costs a
logarithmically divergent energy. In this case, the low-lying elementary
excitations are $S=1$ spin excitations composed of pairs of spinon-vortices.
Similar to the LPP, the spin resonancelike mode is basically composed of a
pair of spinons with $E_{g}=2E_{s}\sim \delta J_{s}$ with $E_{s}=(E_{m})_{%
\mathrm{\min }}$ denoting the lowest level in the spinon spectrum in terms
of $H_{s}$ in Eq. (\ref{hs}), which appears in the dynamic spin
susceptibility at $\mathbf{Q}_{\mathrm{AF}}=(\pi ,\pi )$ as shown in Fig. %
\ref{spinon}(a) in the LPP. However, in the SC phase, the additional
correction from Eq. (\ref{hv}) can modify such a low-lying mode by that for
a pair of spinons with different $m$'s, the confinement force from $F_{h}$
in Eq. (\ref{hv}) can further add a tail in the resonancelike peak from $%
2E_{s}$ towards higher energy.

Another stable elementary excitation in the SC phase is the quasiparticles
composed of spinon-holon pairs as discussed in Ref. \cite{qp2003}. The
d-wave nodal quasiparticle dispersion is schematically shown in Fig. \ref%
{nodalqp}, where the quasiparticle spectral function has a sharp lineshape
below the spinon characteristic energy scale $E_{s}=E_{g}/2$ as
energetically it cannot decay into a spinon and a holon in the condensation
even \emph{locally}. Above $E_{s}$, however, the composite (spin-charge
separation) feature can show up in the lineshape of the spectral function,
since energetically the quasiparticle can decay into a pair of spinon and
holon \emph{locally} without costing much from the logarithmic confining
potential.

Therefore, at low energy and long-distance, there is no more electron
fractionalization in contrast to the LPP. A single spinon-vortex composite
can only emerge as a magnetic vortex core in the mixed state, where it
ensures the minimal flux quantization at $\phi _{0}\equiv hc/2e$ as
discussed in Ref. \cite{muthu2002}. The phase string theory predicts that a
free $S=1/2$ moment must appear in the core of a Arbrikosov magnetic vortex,
even though a Kondo screening effect due to the coupling to the background
quasiparticles may complicate the analysis of possible experimental
observations of such moments.

\begin{figure}[tbp]
\begin{center}
\includegraphics[width=3.5in]{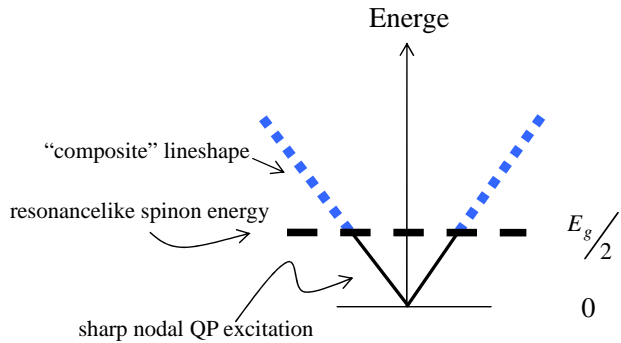}
\end{center}
\caption{{}A schematic nodal quasiparticle dispersion in the
superconducting phase. The lineshape below the characteristic
spinon energy $E_{g}/2$ is very sharp as energetically the
quasiparticle cannot decay into a spinon and a holon in the
condensate. Above $E_{g}/2$, however, such a decay is
energetically allowed \emph{locally} with the spinon and holon
remain loosely confined at large distance. } \label{nodalqp}
\end{figure}

\section{Summary and Conclusions}

In this paper, we have systematically explored the physical properties of
the lower pseudogap phase in the phase string model. In the global phase
diagram of the phase string model, the lower pseudogap phase corresponds to
the regime where the amplitude of the superconducting order parameter
becomes finite, but the order parameter itself is still short of phase
coherence. The presence of free vortices in this phase prevents the system
from undergoing superconducting until some lower temperature is reached
where the vortices and antivortices form bound pairs. So the lower pseudogap
phase and the superconducting phase are intimately related and the former
sets a stage for the latter to emerge.

What makes the present microscopic theory really unique is that the above
picture is naturally embedded in a spin-charge separation description. That
is, for a doped Mott insulator, spin and charge correlations evolve
distinctly, especially at low doping. Thus, at different energy/temperature
scales, the spin and charge degrees of freedom play rather different roles.
For example, at low doping and high temperature the neutral spins start to
form the singlet pairing to lower the superexchange energy, which occurs at
the expense of the kinetic energy of doped holes and the latter behave quite
incoherently in this upper pseudogap phase. So the spin correlations are the
dominant driving force in this high-temperature pseudogap phase, where the
antiferromagnetic fluctuations \emph{continuously} increase with the
decrease of the temperature, and would grow into a true long range order at
zero temperature if there were no intervention from the charge degree of
freedom at some lower temperature at finite doping.

However, the kinetic energy of holes would be highly frustrated in a
long-range antiferromagnetically correlated spin background. Actually the
kinetic energy will eventually win over the long-wavelength spin
correlations beyond a small finite doping concentration at sufficiently low
temperature (Fig. 1). In this regime the trend of growing antiferromagnetic
fluctuations will get stopped and the system makes a crossover into the
lower pseudogap phase at low temperature instead. Here the equal-time spin
correlation length becomes truncated in the same order of magnitude as the
average hole-hole distance [\emph{cf.} Eq.(\ref{xis})]. The corresponding
spin gap behavior in the spin channel is clearly illustrated by the
emergence of a resonancelike peak at energy $E_{g}$ in the spin dynamic
susceptibility at the antiferromagnetic wavevector in Fig. \ref{spinon}(a)
and the suppression of the uniform spin susceptibility at low temperature in
Fig. \ref{spinon}(b).

The spin liquid behavior in the lower pseudogap phase is thus the
consequence of the competition between the charge and spin degrees of
freedom, in which the kinetic energy of doped holes wins. Though suppressed
in density, the thermally activated spin excitations across the spin gap can
still exert frustration effect on the kinetic energy of doped holes as the
reminiscence of the competition between the charge and spin sectors, which
is the cause for the phase disordering. So the superconducting phase can be
finally realized only when those spin excitations which bring destructive
effect on the charge condensate are all excluded out of finite-energy
spectra by some form of confinement.

Technically, in the phase string model represented by the effective
Hamiltonians (\ref{hspinon}) and (\ref{hh}), the spin and charge degrees of
freedom are mathematically characterized by \emph{bosonic} spinon and holon
fields, respectively, which are coupled together by a mutual duality gauge
structure. These \emph{bare} spinons form RVB pairs, and spinon excitations
are the result of the broken RVB pairs. The lower pseudogap phase
corresponds to the condensation of bosonic holons, which obviously is
connected to the kinetic energy gain in the charge sector. The holon
condensation then feedbacks to the spin sector to lead to the suppression of
low-lying antiferromagnetic fluctuations and the opening of the spin gap
through the topological gauge structure. The holon condensation is also
self-consistently strengthened by this spin gap opening effect, because the
fractionalized spin excitations--spinons carry fictitious $\pi $ flux tubes
as perceived by the holons and thus always introduce destructive phase
interference for the latter.

In the lower pseudogap phase, even though the density of states get
suppressed, spinons carrying $\pi $ flux tubes still can be excited
thermally across the energy gap $E_{s}=E_{g}/2$. It is well known that a $%
\pi $ flux tube in a Bose condensate will always induce a vortexlike
supercurrent response. That is precisely what happens in the lower pseudogap
phase and each excited spinon is thus locking into a current vortex to form
the spinon-vortex composite as shown in Fig. \ref{fig_cvortex}. Therefore
the lower pseudogap phase is a vortex liquid/spin liquid phase, which is
also known as the spontaneous vortex phase since these spinon-vortices can
be thermally excited without involving external magnetic field.

So the central prediction of the present microscopic theory is the existence
of spinon-vortices as elementary excitations in the LPP. A spinon-vortex
plays a dual role here. On one hand, it carries a neutral spin-$1/2$ and
indicates the spin-charge separation or electron fractionalization in the
lower pseudogap phase. On the other hand, it is bound to a charge current
vortex and contributes to a $2\pi $ phase vortex in the superconducting
order parameter. The latter property distinguishes the lower pseudogap phase
from the upper pseudogap phase, and represents a mid-step towards the true
spin-charge recombination, which is established in the superconducting phase
where spinon-vortices are \textquotedblleft confined\textquotedblright\ to
form conventional $S=1$ spin excitations and to result in coherent
quasiparticle excitations by the holon-spinon confinement.

The vortex core of a spinon-vortex is distinctly different from a
conventional vortex (like in the BCS superconductor). In the latter a
competing \textquotedblleft normal state\textquotedblright\ is nucleated at
the core, whereas in the former an elementary fractionalized object, spinon,
naturally serves as a core. These spinon-vortices are actually \emph{always}
present, even in the superconducting ground state where all of them are just
paired up, instead of \emph{\textquotedblleft being
annihilated\textquotedblright }. So to create a unpaired vortex at low
energy/temperature, one only needs to break up an RVB pair without
\textquotedblleft nucleating\textquotedblright\ an additional
\textquotedblleft core\textquotedblright . To the leading order of
approximation, the energy scale for creating a vortex is the spinon energy
gap $E_{s}\sim \delta J$ at low doping. In this sense, the spinon-vortices
are really \textquotedblleft cheap\textquotedblright\ vortices which can be
thermally (spontaneously) excited in the lower pseudogap phase. In a similar
sense, the lower pseudogap phase is not a \textquotedblleft
competing\textquotedblright\ phase of essentially different symmetries with
regard to the superconducting state. Rather the former is distinguished from
the latter mainly in the long-wave length limit, much larger than the core
size $a_{c}$ and the RVB pairing size or equal-time spin correlation length,
$\xi _{s}$, where the thermally excited spinon-vortices form
vortex-antivortex pairs to realize the superconducting phase coherence.

Consequently the onset of the LPP at $T_{v}$ or $H_{v}$, though a charge
behavior, is determined by the spinon excitation according to Eq. (\ref{tv}%
), which can be understood based on the \textquotedblleft core
touching\textquotedblright\ picture of spinon vortices with the
\textquotedblleft supercurrents\textquotedblright\ being totally squeezed
out. The Nernst effect has been also connected to the spin degrees of
freedom as the latter provide transport entropy, while the diamagnetism in
the low field limit is associated with the configurational entropy of
spinon-vortices. The overall theoretical results are found to be in the
correct magnitude as compared with the experimental measurements in the
cuprates. Furthermore, a conserved, dissipationless spin Hall effect has
been also predicated as a unique feature for the LPP. We shall present more
detailed and quantitative results based on the mutual Chern-Simons gauge
theory of the phase string model in a separate publication\cite{qi2006}.

\begin{acknowledgments}
We would like to thank V. N. Muthukumar, S. P. Kou, Z. C. Gu, and W. Q. Chen
for stimulating discussions and early collaborations. We also thank P. W.
Anderson, N. P. Ong, D. N. Sheng, Y. Y. Wang, H. H. Wen, and F. C. Zhang for
very helpful discussions. The authors acknowledge the support of the grants
from the NSFC and MOE.
\end{acknowledgments}

\bibliographystyle{plain}
\bibliography{hightc}

\begin{thebibliography}{99}
\bibitem{remark1} One may also directly use the GL equation (\ref{gl0}) to
get a similar approximate estimate for $T_{v}$, by rewriting it in the form $%
(-i\nabla -\mathbf{A}^{s})^{2}\psi _{h}=(1/\xi _{h}^{2})(1-|\psi
_{h}|^{2}/\rho _{h}^{0})\psi _{h}$ and by noting that $\xi _{h}\equiv 1/%
\sqrt{2m_{h}(-\alpha )}>$ the average holon-holon inter-distance $\sim a_{c}$
while $(-i\nabla -\mathbf{A}^{s})^{2}\psi _{h}\sim \left( 1/a_{c}^{2}\right)
\psi _{h}$ in the vortex packing limit.
\end{thebibliography}

\end{document}